\documentclass[aps,pre,preprint, showkeys, superscriptaddress]{revtex4}
\pdfoutput=1 
\usepackage{latexsym}
\usepackage{natbib}
\usepackage{amsmath}
\usepackage{amsfonts}
\usepackage{amssymb}
\usepackage[T1]{fontenc}
\usepackage{pdfpages}
\usepackage{epstopdf}
\usepackage{graphicx}  
\usepackage{verbatim}   
\usepackage{t1enc}
\usepackage{anysize}
\usepackage{tocbibind}
\usepackage{subfig}
\usepackage[percent]{overpic} 
\usepackage{float}  
\usepackage{appendix} 
\usepackage{booktabs} 
\usepackage{natbib}
\usepackage{multirow}
\usepackage{array}

\DeclareGraphicsExtensions{.pdf,.eps}

\begin{document}
\title{Failure-recovery model with competition between failures in complex networks: a dynamical approach}
\author{L. D. Valdez} \affiliation{Instituto de Investigaciones
  F\'isicas de Mar del Plata (IFIMAR)-Departamento de F\'isica,
  Facultad de Ciencias Exactas y Naturales, Universidad Nacional de
  Mar del Plata-CONICET, Funes 3350, (7600) Mar del Plata, Argentina.}
\author{M. A. Di Muro} \affiliation{Instituto de Investigaciones
  F\'isicas de Mar del Plata (IFIMAR)-Departamento de F\'isica,
  Facultad de Ciencias Exactas y Naturales, Universidad Nacional de
  Mar del Plata-CONICET, Funes 3350, (7600) Mar del Plata, Argentina.}
\author{L. A. Braunstein} \affiliation{Instituto de Investigaciones
  F\'isicas de Mar del Plata (IFIMAR)-Departamento de F\'isica,
  Facultad de Ciencias Exactas y Naturales, Universidad Nacional de
  Mar del Plata-CONICET, Funes 3350, (7600) Mar del Plata, Argentina.}
\affiliation{Center for Polymer Studies, Boston University, Boston,
  Massachusetts 02215, USA}

\begin{abstract}
Real systems are usually composed by units or nodes whose activity can
be interrupted and restored intermittently due to complex interactions
not only with the environment, but also with the same
system. Majdand\v{z}i\'c $et\;al.$ [Nature Physics {\bf 10}, 34
  (2014)] proposed a model to study systems in which active nodes fail
and recover spontaneously in a complex network and found that in the
steady state the density of active nodes can exhibit an abrupt
transition and hysteresis depending on the values of the
parameters. Here we investigate a model of recovery-failure from a
dynamical point of view. Using an effective degree approach we find
that the systems can exhibit a temporal sharp decrease in the fraction
of active nodes. Moreover we show that, depending on the values of the
parameters, the fraction of active nodes has an oscillatory regime
which we explain as a competition between different failure
processes. We also find that in the non-oscillatory regime, the
critical fraction of active nodes presents a discontinuous drop which
can be related to a ``targeted'' k-core percolation process. Finally, using mean
field equations we analyze the space of parameters at which hysteresis
and oscillatory regimes can be found.
\end{abstract}


\maketitle
\section{Introduction}

In nature and social networks, node aging effects and external forces
introduce perturbations on these systems which affect their functions
or even can trigger catastrophic cascade of failures. However many of
these systems are able to develop different mechanisms to recover
their functionality. For instance, it was recently shown that rat
brains under anesthesia pass through discrete metastable states of
activity which allows to recover from a state of induced comma to a
full consciousness state in a physiological time~\cite{Hud_01}.
In protein network regulation when, for example the DNA is damaged, a
specific protein is activated~\cite{Bos_01}. This produces the arrest
of the cell division cycle which prevents the proliferation of cells
containing damaged DNA (tumor formation). Then, a biochemical
processes involved in DNA repair is initiated. Once this task is
completed successfully, the cell resumes its progression so that cell
division can take place. If repairing is not possible due to excessive
damage, the specific protein leads to apoptosis, $i.e.$ programmed cell
death.

Recently Majdand\v{z}i\'c $et\; al.$~\cite{Maj_01} proposed a model to
study systems in which nodes fail and recover spontaneously in a
complex network. In their model, a node can be in one of the following
two states: active or inactive. In particular, nodes can be inactive
due to: i) internal failure (independently of the states of their
neighbors) or ii) external failure when a fraction of their neighbors
are inactive, $i.e.$ there is an interaction between nodes and their
neighbors. They studied numerically and theoretically, in a mean field
approach, the steady state of the process and found that the density
of active nodes $A$ can exhibits an abrupt transition and hysteresis,
which mimics the behavior observed in different biological and
economical systems~\cite{Maj_01}.

The model proposed by Majdand\v{z}i\'c $et\; al.$~\cite{Maj_01} can be related to an
epidemic model, since active nodes are equivalent to susceptible
nodes, $i.e.$, non-infected individuals; and inactive nodes are
equivalent to infected ones. As a consequence, the same tools
implemented in the field of epidemiology can be extended to models
where nodes recover and fail spontaneously as in Ref.~\cite{Maj_01}.

In this manuscript we propose a dynamical model of activation and
spontaneous recovery and use the framework from the epidemiology field
to describe the dynamics of the process. The study of dynamical
processes in complex systems is a very important area of research
since it allows understanding the role of the nonlinearities involved
in the processes. There are different theoretical approaches to study
the evolution of a disease spreading. One of the most detailed
framework is the Markovian equations applied to complex networks, to
study the evolution of the spread of an epidemic~\cite{Van_01}. In
this approach, it is necessary to use an order of $N$ differential
equations, where $N$ is the size of the network. Another theoretical
tool is the effective degree approach, \cite{Lin_01} in which the
compartments are disaggregated by the states of the nodes of a network
(infected or non-infected), and by the number of its neighbors in each
state. In particular, in epidemic models such as the
Susceptible-Infected-Recovered (SIR) and the
Susceptible-Infected-Susceptible (SIS) ---see Ref.~\cite{Lin_01}---
the number of equations used to describe the evolution of the density
of individuals in different compartments is of the order
$O(k_{\text{max}}^3)$ and $O(k_{\text{max}}^2)$ respectively, where
$k_{\text{max}}$ is the maximum degree that a node can have. In
Ref.~\cite{Lin_01} it was shown that this approach gives a good
agreement between theory and simulations for the SIR and SIS
models. Finally, one of the simplest tools to study epidemic
process are the equations based on the law of mass action, or simply,
mean field (MF) equations~\cite{And_01} which have very little or no
information about the topology of the network and disregard any
correlation between the states of the nodes.  Although sometimes,
there is not a good agreement between the theoretical results and the
simulations on complex networks, this approach: i) gives a
qualitatively description of the process, ii) allows to find
analytically the behavior of relevant magnitudes, iii) allows to study
the stability of the fixed points in MF easily. For interested readers
a more detailed description of the tools applied on epidemic models
can be found in~\cite{Pas_01,Tay_01,Mil_01,Roc_01,Gle_01} and
references therein.

In this work we apply the degree based framework, used in epidemic
processes that spread in complex networks, to describe the evolution
of the states on complex networks where active nodes overcome
internal, external failures and recovery. In our failure-recovery
model active nodes can fail by random internal failures at a rate $p$
and recover from this kind of failure at a rate $\gamma_I$. Active
nodes can also fail at a rate $r$ due to lack of support of their
neighborhood and recover at a rate $\gamma_E$. Unlike the model
presented in Ref~\cite{Maj_01}, here we distinguish between inactive
nodes failed by internal and external failures that dynamically
compete to ``capture'' active nodes . Our model mimics some biological
systems such as neural networks, where some nodes can exhibit
inhibitory or excitatory functions~\cite{Gol_01}.

We find that depending on the values of the parameters, the system
exhibits regimes with hysteresis and oscillations. We discuss the
relation between our model in the steady state and k-core
percolation. Finally, using a MF approach we study the phase diagram
of the fraction of active nodes as a function of the parameters, and
we show that only for $\gamma_I<\gamma_E$ the system is able to sustain
oscillations.

This paper is organized as following: in Sec.~\ref{Sec.mod} we present
our model and the evolution equations based on the effective degree
approach. In Sec.~\ref{Sec.ResTime} we show our dynamical results and in
Sec.~\ref{Sec.SteStaEff} we present the results in the steady
state. In Sec.~\ref{Sec.StaAn} we study the stability of the solutions
and construct the Lyapunov function in the mean field approach. In
Sec.~\ref{Concl} we present our conclusions.

\section{Model}\label{Sec.mod}
\begin{figure}[H]
\centering
\vspace{0.5cm}
  \begin{overpic}[scale=0.30]{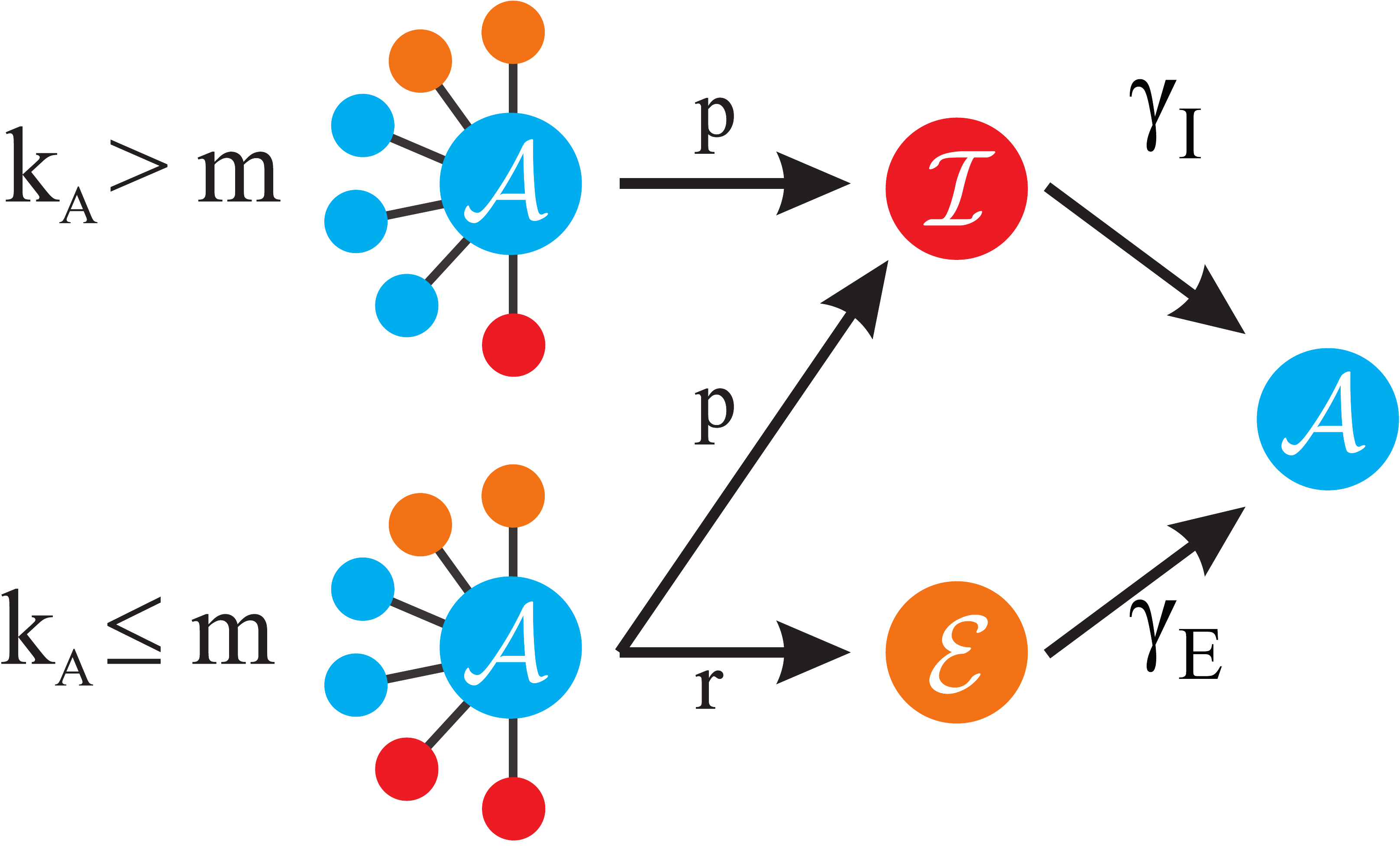}
    \put(15,50){}
  \end{overpic}\hspace{.50cm}
\caption{Schematic representation of the model. Light blue, red and
  orange nodes represent the active ($\mathcal{A}$), inactive due to
  internal failure ($\mathcal{I}$) and inactive due to external
  failure ($\mathcal{E}$), respectively. Active nodes can fail
  internally with rate $p$ independently of the number of active
  neighbors ($k_A$). Active nodes with $k_A\leq m$ can also fail
  externally with rate $r$ due to the lack of support of their
  neighbors. In the schematic we use $m=2$. The $\mathcal{E}$ nodes
  become active at rate $\gamma_E$ and the $\mathcal{I}$ ones at rate
  $\gamma_I$.}\label{fig.esqRules}
\end{figure}

In our failure-recovery model, a node can be in one of the following
three compartment states:
\begin{itemize}
\item Active ($\mathcal{A}$): nodes which are not failed or damaged,
\item Inactive due to internal failure ($\mathcal{I}$): $\mathcal{A}$
  nodes that fail at a rate $p$ independently of the states of their
  neighbors. These nodes recover ($i.e$ become active) at a rate
  $\gamma_I$,
\item Inactive due to external failure ($\mathcal{E}$): $\mathcal{A}$
  nodes having $m$ or less active neighbors which fail at a rate $r$
  due to lack of support from their neighbors. These nodes recover at
  a rate $\gamma_E$.
\end{itemize}
In Fig.~\ref{fig.esqRules} we show a schematic of the rules of our
spontaneous recovery model.
The particular case of $\gamma_I=\gamma_E=p=0$ is the special case of
a k-core percolation process~\cite{Dor_01,Cel_01,Cel_02,Bax_02}
in which nodes go through an irreversible transition from state
$\mathcal{A}$ to $\mathcal{E}$. In the ``random'' k-core percolation,
after randomly removing a fraction $1-q$ of nodes, a cascade is
triggered and all the nodes having $m$ or less non-removed or living
neighbors, are removed. In the steady state, there is a giant
component (GC) composed by nodes with more than $m$ living neighbors,
which we call a ``compact'' sub-graph. It was shown that the final
number of living nodes in this process can exhibit a first order phase
transition at a critical initial failure $1-q_{c}$ where $q_{c}$, is
the initial critical fraction of living nodes in the cascade. In
Appendix~\ref{Sec.Akcore} we show the equations for the steady state
of the ``random'' k-core percolation. We will also discuss later the
relation between our model and k-core percolation.

The main theoretical approach that we use in this manuscript to
describe our model is the effective degree approach~\cite{Lin_01} that will
be compared with the stochastic simulations. However, in order to
study qualitatively the phase diagrams and the stability of the
solutions, we will use mean field equations obtained from the degree
based approach in which the correlations between the states of
nodes and their neighbors are disregarded.

\subsection{Effective degree approach and mean field equations}
For the effective degree approach, first introduced by Lindsquit
$et\;al.$~\cite{Lin_01}, the compartments are disaggregated by the states of
the nodes of a network ($\mathcal{A}$, $\mathcal{I}$, $\mathcal{E}$),
and by the number of its neighbors in each state. We denote by
$A(k_A,k_I,k_E)$ [and similarly $I(k_A,k_I,k_E)$ and $E(k_A,k_I,k_E)$]
the density of active nodes (internal inactive and external inactive)
with $k_A$, $k_I$ and $k_E$ neighbors in state $\mathcal{A}$,
$\mathcal{I}$ and $\mathcal{E}$, respectively; where $k_A+k_I+k_E=k$
is the degree of a node. In our model, the flow into and outside these
compartments are due to the change on the state of the nodes and their
neighbors. The evolution equations for the states in our
failure-recovery model are given by
\begin{eqnarray}
\frac{dA(k_A,k_I,k_E)}{dt}&=&\gamma_II(k_A,k_I,k_E)+\gamma_EE(k_A,k_I,k_E)+\nonumber\\
                          &&-rA(k_A,k_I,k_E)\Theta(m-k_A)-p A(k_A,k_I,k_E)+\nonumber\\
                          &&\gamma_E [(k_E+1)A(k_A-1,k_I,k_E+1)-k_EA(k_A,k_I,k_E)]+ \nonumber \\
                          &&\gamma_I[(k_I+1)A(k_A-1,k_I+1,k_E)-k_IA(k_A,k_I,k_E)]+\nonumber\\
                          &&p [(k_A+1)A(k_A+1,k_I-1,k_E)-k_AA(k_A,k_I,k_E)]+ \nonumber\\
                          && rW_A[(k_A+1)A(k_A+1,k_I,k_E-1)-k_AA(k_A,k_I,k_E)],\label{eq.lind1}\\
\frac{dI(k_A,k_I,k_E)}{dt}&=&-\gamma_I I(k_A,k_I,k_E) +p A(k_A,k_I,k_E) + \nonumber\\
                          &&\gamma_E [(k_E+1)I(k_A-1,k_I,k_E+1)-k_E I(k_A,k_I,k_E)]+ \nonumber\\
                          &&\gamma_I[(k_I+1)I(k_A-1,k_I+1,k_E)-k_I I(k_A,k_I,k_E)] + \nonumber\\
                          &&p [(k_A+1)I(k_A+1,k_I-1,k_E) - k_A I(k_A,k_I,k_E)]+\nonumber\\
                          &&+r W_I [(k_A+1)I(k_A+1,k_I,k_E-1)-k_AI(k_A,k_I,k_E)],\label{eq.lind2}\\
\frac{dE(k_A,k_I,k_E)}{dt}&=& r A(k_A,k_I,k_E) \Theta(m-k_A) -\gamma_E E(k_A,k_I,k_E)+\nonumber\\
                          &&\gamma_E[(k_E+1)E(k_A-1,k_I,k_E+1)-k_E E(k_A,k_I,k_E)]+\nonumber\\
                          &&+\gamma_I[(k_I+1)E(k_A-1,k_I+1,k_E)-k_I E(k_A,k_I,k_E)]+\nonumber\\
                          &&p [(k_A+1)E(k_A+1,k_I-1,k_E)-k_A E(k_A,k_I,k_E)]+\nonumber\\
                          &&r W_E [(k_A+1)E(k_A+1,k_I,k_E-1)-k_AE(k_A,k_I,k_E)],\label{eq.lind3}
\end{eqnarray}
where $\Theta(x)$ is the Heaviside distribution. In these equations
$k_{\text{min}}\leq k_A+k_I+k_E \leq k_{\text{max}}$, where
$k_{\text{min}}$ and $k_{\text{max}}$ are the maximum and minimum
degree of the degree distribution $P(k)$. Here, $P(k)$ represents the
fraction of nodes with $k$ neighbors, $i.e.$ with degree
$k$. 

Eq.~(\ref{eq.lind1}) [and similarly Eqs.~(\ref{eq.lind2}) and
  (\ref{eq.lind3})] represents the evolution of the density of active
nodes ($I$ and $E$) with $k_A$, $k_E$ and $k_I$ neighbors in states
$\mathcal{A}$, $\mathcal{I}$, $\mathcal{E}$, respectively [or with
  neighborhood $(k_A,k_E,k_I)$]. Notice that the information of the
degree distribution $P(k)$ is encoded in the initial condition of the
system of Eqs.~(\ref{eq.lind1})-(\ref{eq.lind3}). For example, for the
initial condition in which all nodes are active,
$A(k_A=k,k_I=0,k_E=0)=P(k)$.

\begin{figure}[H]
\centering
\vspace{0.5cm}
  \begin{overpic}[scale=0.30]{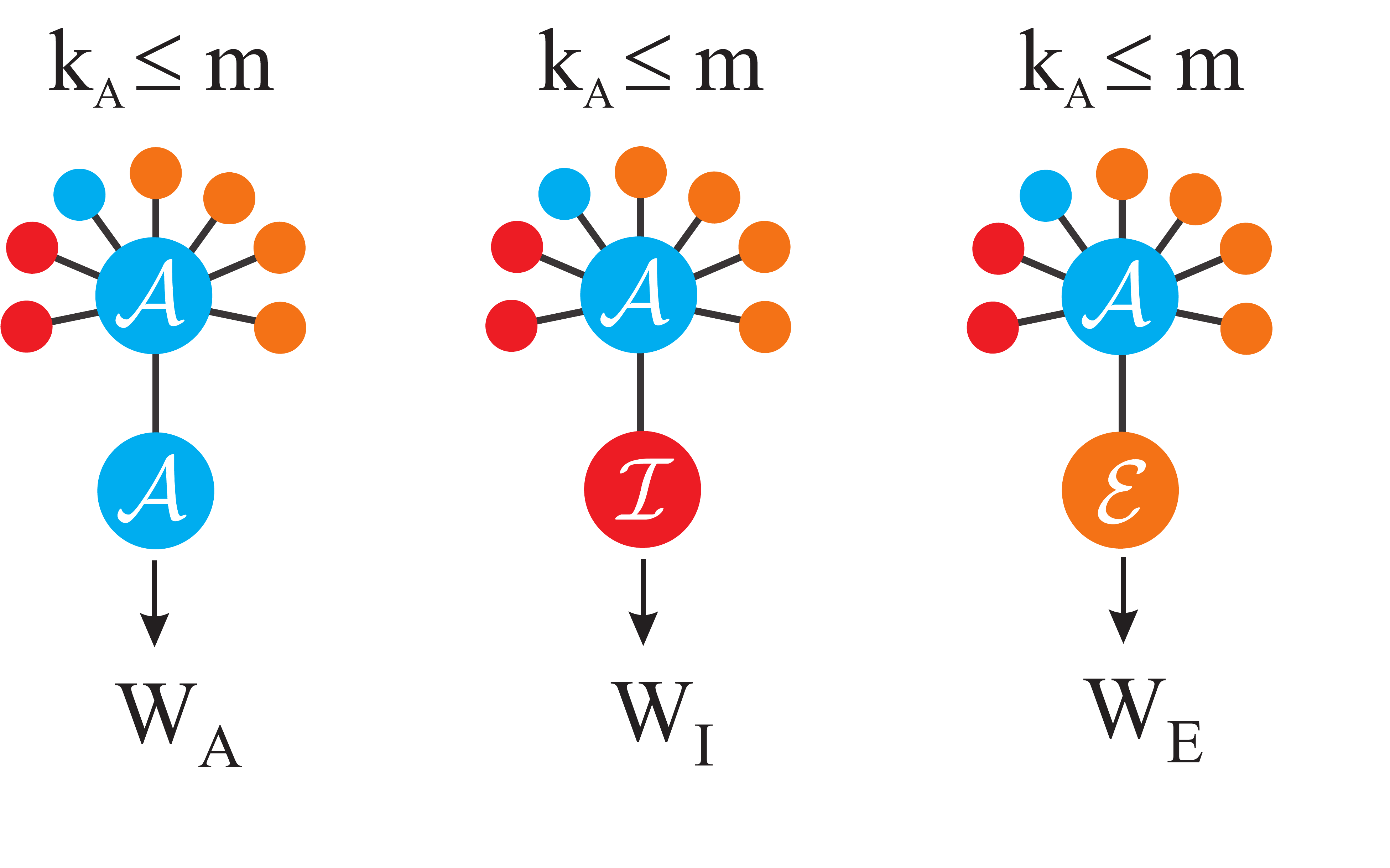}
    \put(-5,60){(a)}
    \put(30,60){(b)}
    \put(65,60){(c)}
  \end{overpic}\hspace{.50cm}
\caption{Schematic representation of the terms $W_A$ (a), $W_I$ (b)
  and $W_E$ (c) for $m=2$. The colors of the nodes represent the same
  as in Fig.~\ref{fig.esqRules}. $W_A$ represents the fraction of
  edges connecting two active nodes, in which one of them has $k_A
  \leq m$. Similarly, $W_I$ ($W_E$) represents the fraction of edges
  connecting nodes in state $\mathcal{I}$ ($\mathcal{E}$) with nodes
  in state $\mathcal{A}$ with $k_A \leq m$.}\label{fig.W}
\end{figure}

In the r.h.s Eq.~(\ref{eq.lind1}) the term:
\begin{itemize}
\item $\gamma_I\;I(k_A,k_I,k_E)$ represents the transition from a node
  in state $\mathcal{I}$ with a neighborhood $(k_A,k_I,k_E)$, to state
  $\mathcal{A}$, due to the recovery of these inactive
  nodes,
\item $\gamma_E\;E(k_A,k_I,k_E)$ corresponds to the transition from
  state $\mathcal{E}$ to $\mathcal{A}$ due to recovery at rate
  $\gamma_E$,
\item $r\;A(k_A,k_I,k_E)\Theta(m-k_A)$ depicts the density of active
  nodes with $k_A\leq m$ that fail externally at a rate $r$,
\item $p\;A(k_A,k_I,k_E)$ represents the transition from nodes with
  state $\mathcal{A}$ and neighborhood $(k_A,k_I,k_E)$, to nodes with
  state $\mathcal{I}$ at a rate $p$ due to internal failure,
\item $p [(k_A+1)A(k_A+1,k_I-1,k_E) - k_A A(k_A,k_I,k_E)]$ represents
  the transition, in which neighbors in state $\mathcal{A}$ becomes
  $\mathcal{I}$,
\item $\gamma_E [(k_E+1)A(k_A-1,k_I,k_E+1)-k_EA(k_A,k_I,k_E)]$ is the
  transition in which neighbors in state $\mathcal{E}$ become active
  at a rate $\gamma_E$,
\item $\gamma_I[(k_I+1)A(k_A-1,k_I+1,k_E)-k_I
  A(k_A,k_I,k_E)] $ represents the transition from neighbors in
  state $\mathcal{I}$ to $\mathcal{A}$ at a rate $\gamma_I$, and finally,
\item $r\;W_A [(k_A+1)A(k_A+1,k_I,k_E-1)-k_AA(k_A,k_I,k_E)]$ represents
  the density of active nodes whose neighbors in state $\mathcal{A}$
  become $\mathcal{E}$.
\end{itemize}
Here $W_A$, ($W_I$ and $W_E$) represents the probability that an
active neighbor (with $k_A \leq m$) is connected to a node in state
$\mathcal{A}$, ($\mathcal{I}$ and $\mathcal{E}$) (see
Fig.~\ref{fig.W}). Notice that the last four terms depict the
transitions of a node caused by its neighbors and not by changes in
its own state.

In Table~\ref{tab.Trans} we show the flow into and outside the
compartment $A(k_A,k_I,k_E)$  [see Eqs.~(\ref{eq.lind1})].

\begin{table}[H]
\centering
\caption{Transitions involved in Eq.~(\ref{eq.lind1}).}
\label{tab.Trans}
\begin{tabular}{|c|c|}
\hline
Transition & Rate  \\
\hline
$A(k_A,k_I,k_E) \to E(k_A,k_I,k_E)$ & $-r$ \\
$A(k_A,k_I,k_E) \to I(k_A,k_I,k_E)$ & $-p$ \\
$A(k_A,k_I,k_E) \to A(k_A+1,k_I,k_E-1)$ &  $-\gamma_E\;k_E$ \\
$A(k_A,k_I,k_E) \to A(k_A+1,k_I-1,k_E)$ & $-\gamma_I\;k_I$ \\
$A(k_A,k_I,k_E) \to A(k_A-1,k_I+1,k_E)$ & $-p\;k_A$ \\
$A(k_A,k_I,k_E) \to A(k_A-1,k_I,k_E+1)$ &  $-r\;W_A k_A$\\
$I(k_A,k_I,k_E) \to A(k_A,k_I,k_E)$ & $\gamma_I$  \\
$E(k_A,k_I,k_E) \to A(k_A,k_I,k_E)$ & $\gamma_E$ \\
$A(k_A-1,k_I,k_E+1) \to A(k_A,k_I,k_E)$ & $\gamma_E\;(k_E+1)$  \\
$A(k_A-1,k_I+1,k_E) \to A(k_A,k_I,k_E)$ &  $\gamma_I\;(k_I+1)$ \\
$A(k_A+1,k_I-1,k_E)\to A(k_A,k_I,k_E)$ &  $p\;(k_A+1)$\\
$A(k_A+1,k_I,k_E-1) \to A(k_A,k_I,k_E)$ & $r\;W_A\;(k_A+1)$ \\
\hline
\end{tabular}
\end{table}

It is straightforward the interpretation of each term of
Eqs.~(\ref{eq.lind2}) and (\ref{eq.lind3}).  Note that the last
relation in Table~\ref{tab.Trans} represents an effective dynamical
rate of transition at which active neighbors of an active node fail
externally, which is proportional to $W_A$, that is, the ratio between
the mean number of active neighbors of an active node that can fail
and the total mean number of active neighbors:
\begin{eqnarray}
W_A=\frac{\sum_{k_A=0}^{m}\sum_{k_I=0}^{k_{\text{max}}}\sum_{k_E=0}^{k_{\text{max}}}k_AA(k_A,k_I,k_E)}{\sum_{k_A=0}^{k_{\text{max}}}\sum_{k_I=0}^{k_{\text{max}}}\sum_{k_E=0}^{k_{\text{max}}}k_AA(k_A,k_I,k_E)}.\nonumber\\
\end{eqnarray}
Similarly $W_I$, $W_E$ are given by
\begin{eqnarray}
W_I=\frac{\sum_{k_A=0}^{m}\sum_{k_I=0}^{k_{\text{max}}}\sum_{k_E=0}^{k_{\text{max}}}k_IA(k_A,k_I,k_E)}{\sum_{k_A=0}^{k_{\text{max}}}\sum_{k_I=0}^{k_{\text{max}}}\sum_{k_E=0}^{k_{\text{max}}}k_IA(k_A,k_I,k_E)},\nonumber\\
W_E=\frac{\sum_{k_A=0}^{m}\sum_{k_I=0}^{k_{\text{max}}}\sum_{k_E=0}^{k_{\text{max}}}k_EA(k_A,k_I,k_E)}{\sum_{k_A=0}^{k_{\text{max}}}\sum_{k_I=0}^{k_{\text{max}}}\sum_{k_E=0}^{k_{\text{max}}}k_EA(k_A,k_I,k_E)}.\nonumber
\end{eqnarray}

From the system of Eqs.~(\ref{eq.lind1})-(\ref{eq.lind3}) the density
of nodes in states $\mathcal{A}$, $\mathcal{I}$ and $\mathcal{E}$,
that we denote by $A$, $I$ and $E$ respectively, are
given by,

\begin{eqnarray}
 A  &\equiv &\sum_{k_A=0}^{k_{\text{max}}}\sum_{k_I=0}^{k_{\text{max}}}\sum_{k_E=0}^{k_{\text{max}}}A(k_A,k_I,k_E),\label{eq.Amed}\\
 I  & \equiv &\sum_{k_A=0}^{k_{\text{max}}}\sum_{k_I=0}^{k_{\text{max}}}\sum_{k_E=0}^{k_{\text{max}}}I(k_A,k_I,k_E),\\
 E  &\equiv&\sum_{k_A=0}^{k_{\text{max}}}\sum_{k_I=0}^{k_{\text{max}}}\sum_{k_E=0}^{k_{\text{max}}}E(k_A,k_I,k_E).
\end{eqnarray}

The agreement between Eqs.~(\ref{eq.lind1})-(\ref{eq.lind3}) and the
simulations improves as the mean connectivity $\langle k \rangle=\sum k P(k)$
increases. Therefore, in order to compare the effective degree
equations with the stochastic model, in the following sections we
present the results based on Random Regular (RR) networks, where all the
nodes have the same degree $z=32$, and in non-regular networks
constructed using the Configurational Model~\cite{Mol_01} with
$\langle k \rangle =32$. For networks with a smaller mean connectivity
($\langle k \rangle\approx 10$) we show only the simulations. In the
stochastic model we use $N=10^6$ and the Gillespie's algorithm.

\section{Results}\label{Sec.Res}
\subsection{Time evolution}\label{Sec.ResTime}

We compute the density of nodes in state $\mathcal{A}$, $\mathcal{I}$
and $\mathcal{E}$ in the steady state of our failure-recovery model as
a function of $p^{*}=1-\exp(-p/\gamma_I)$ (see Ref.~\cite{Maj_01}),
which is a convenient parameter to show our numerical results since
$p^{*}\in [0,1]$ ($p^{*}=0$ for $p=0$ and $p^{*}=1$ for
$p=\infty$). Additionally, for small values of $p$ ($p << \gamma_I$),
$p^*$ corresponds to the steady density of nodes in state
$\mathcal{I}$ of our model when $r=0$, $i.e.$ when there is no state
$\mathcal{E}$, and nodes become $\mathcal{A}$ and $\mathcal{I}$
intermittently without any interaction between them (see
Appendix~\ref{appPstar}).

In Fig.~\ref{fig.TempS10}, we show the evolution of the density of
active nodes [see Eq.~(\ref{eq.Amed})] for RR network, obtained from
the simulation and from Eqs.~(\ref{eq.lind1})-(\ref{eq.lind3}) for
$\gamma_I=0.01$, $\gamma_E=1$, $r=5$, $p^{*}=0.40$ and $m=8$ for two
different initial conditions, $A=1$ in (a) and $I=1$ in (b). Notice
that $A+E+I=1$.

\begin{figure}[H]
\centering
\vspace{0.5cm}
  \begin{overpic}[scale=0.30]{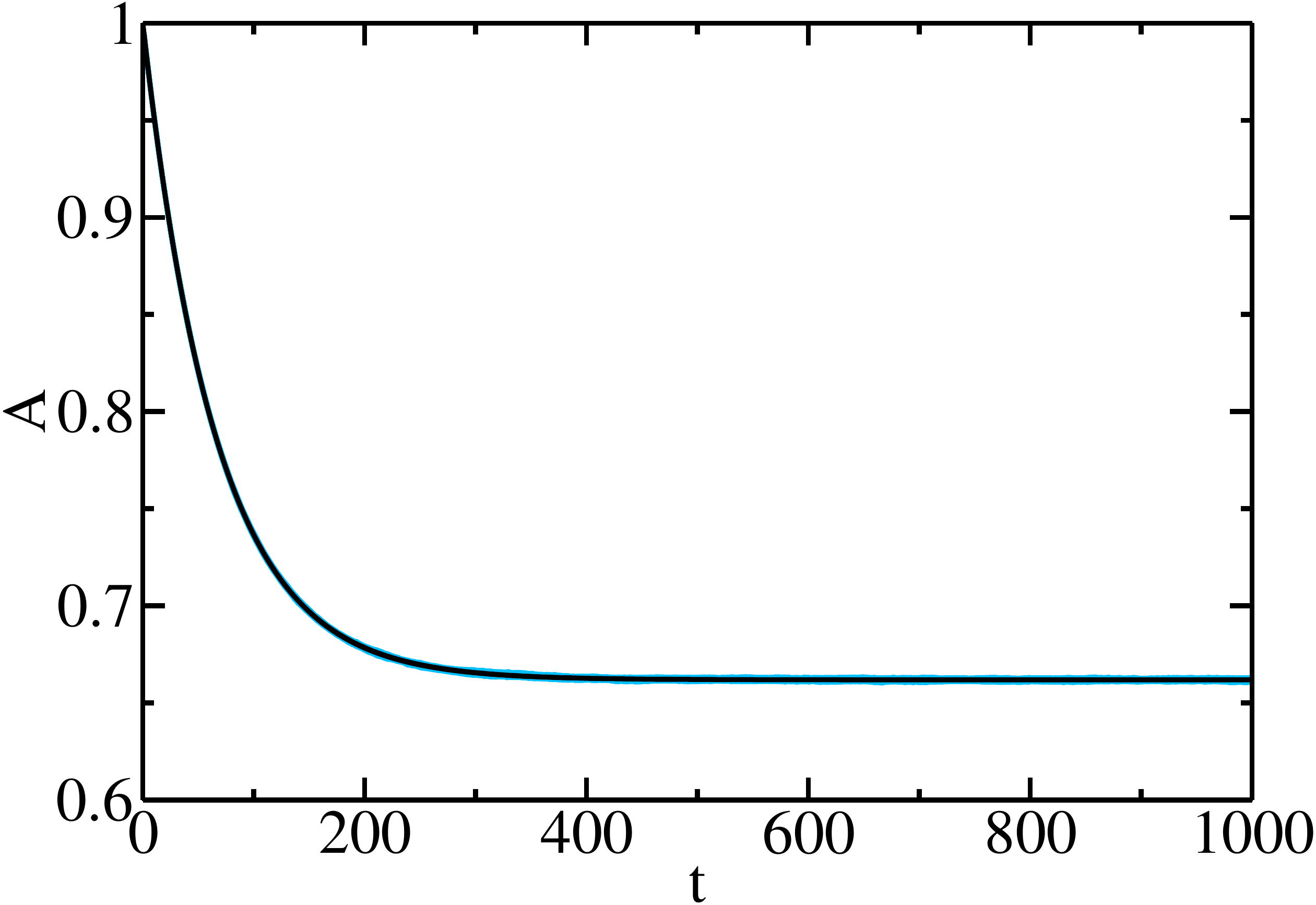}
    \put(15,50){{\bf{(a)}}}
  \end{overpic}\hspace{.50cm}
  \begin{overpic}[scale=0.30]{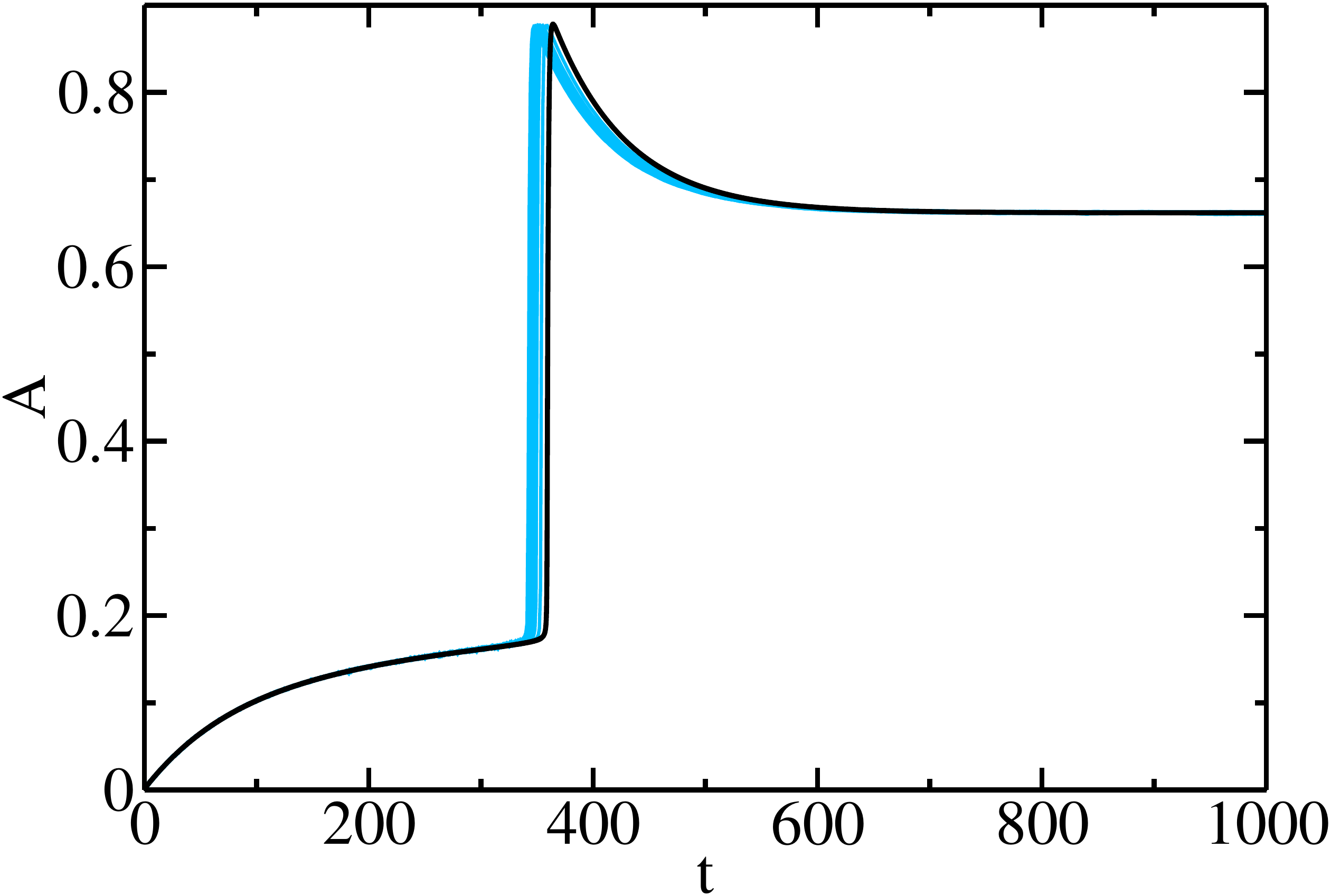}
    \put(15,50){{\bf{(b)}}}
  \end{overpic}\hspace{0.5cm}
\caption{Temporal evolution of the density of active nodes for RR
  networks with $z=32$ for $\gamma_I=0.01$, $\gamma_E=1$, $r=5$,
  $m=8$, $p^{*}=0.40$ for two initial conditions: (a) $A=1$ and (b)
  $I=1$. The theoretical solutions (black) are obtained from the
  degree effective equations~(\ref{eq.lind1})-(\ref{eq.lind3}) and the
  simulations results (colored lines) are the results of $100$
  different network realizations with $N=10^6$.}\label{fig.TempS10}
\end{figure}

From Fig.~\ref{fig.TempS10} we can see that the theoretical model is
in well agreement with the simulation. In Fig.~\ref{fig.TempS10} (b)
we can see that there is a slightly difference between the simulations
and the theory on the time at which the density of active nodes rises
sharply. This difference can be explained by stochastic effects
similarly than the one found in epidemic models~\cite{Bar_01} when the
initial condition consists in a few infected nodes. For this case, the
time at which the density of infected individuals grows sharply varies
for different realizations~\cite{Bar_01}. For the parameters used in
Fig.~\ref{fig.TempS10}, the system reaches a steady state, however, we
will show that for a specific region of parameters the system exhibits
an oscillatory behavior, similarly to the ones found in some epidemic
models~\cite{Roz_01,Roz_02,Kup_01} and in a model of neural
networks~\cite{Gol_01}. In our model, these oscillations are a
consequence of a competition between inactive internal nodes and
inactive external nodes, with the aim of transforming the living nodes
to their own state, as we will explain below.

\begin{figure}[H]
\centering
\vspace{0.5cm}
  \begin{overpic}[scale=0.30]{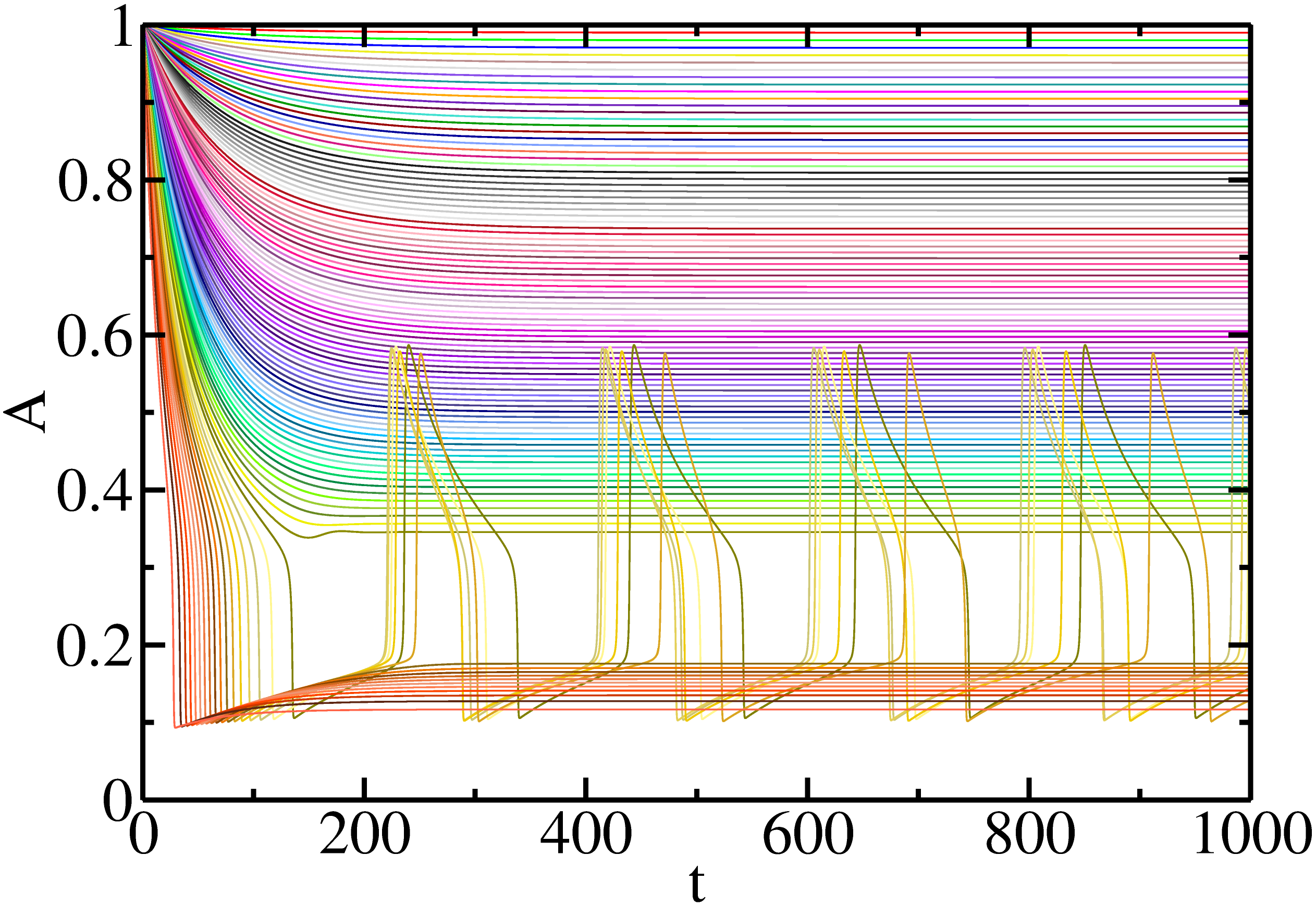}
    \put(15,70){{\bf{(a)}}}
  \end{overpic}\hspace{0cm}
  \begin{overpic}[scale=0.30]{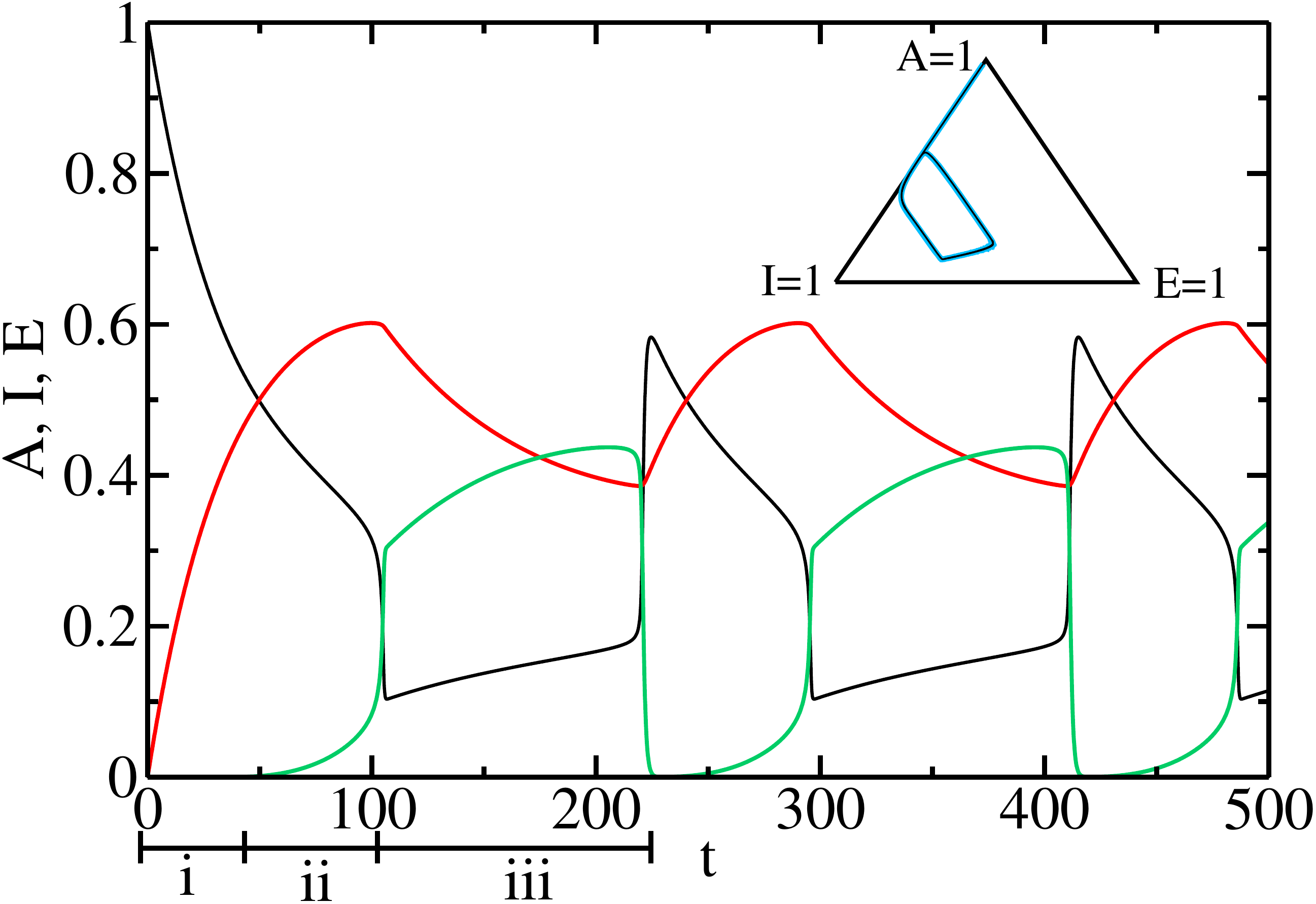}
    \put(10,70){{\bf{(b)}}}
  \end{overpic}\hspace{0cm}
\caption{For RR networks with $\gamma_I=0.01$, $\gamma_E=1$, $r=3$,
  $m=8$: (a) density of active nodes as a function of time obtained
  from the effective degree approach for $p^{*}=1-\exp(-p/\gamma_I)$
  from $0.00$ (top) to $0.99$ (bottom) with $\delta p^*=10^{-2}$; and
  (b): temporal evolution of the density $A$ (black), $I$ (red) and
  $E$ (green) with $p=0.01897$ ($p^*=0.85$) obtained from the
  effective degree approach. The intervals of time $i$, $ii$ and $iii$
  correspond qualitatively to different regimes due to the competition
  between nodes in state $\mathcal{E}$ and $\mathcal{I}$ (explained in
  the text). In the inset we show the phase portrait in a triangular
  simplex obtained from the main plot (black) and we compare the
  results with four stochastic network realizations (colored lines)
  with $N=10^6$ nodes.}\label{fig.ERSFesq}
\end{figure}

In Fig.~\ref{fig.ERSFesq} (a), we plot the theoretical results for the
evolution of the density of active nodes for the same parameters of
Fig.~\ref{fig.TempS10}(a), but for $r=3$, instead of the value $r=5$
used in Fig.~\ref{fig.TempS10} for different values of $p^{*}$. From
the figure we can see that the system exhibits oscillatory behavior in
the range $0.83 \leq p^{*} \leq 0.88$.

The oscillatory phase can be explained as a competition between internal and external inactive nodes to turn active nodes
into states $\mathcal{I}$ and $\mathcal{E}$, respectively. The dynamic
of this competition is shown in Fig.~\ref{fig.ERSFesq} (b) in which we
identify qualitatively three consecutive regimes ($i$, $ii$ and
$iii$):
\begin{itemize}
\item ($i$): Initially, in this interval all nodes are active and they
  can only fail internally because each active node has $k_A>m$
  neighbors. Therefore $A$ goes down and $I$ rises while $E$ remains
  near to zero.
\item ($ii$): In this stage as $I$ increases, the fraction of nodes in
  state $\mathcal{E}$ raises faster than in the previous stage since
  there is an increasing number of active nodes with $k_A \leq m$
  neighbors. As these nodes in state $\mathcal{A}$ become
  $\mathcal{E}$, there are less available active nodes that can make a
  transition to $\mathcal{I}$ states, which is reflected in a slower
  increasing of $ I$, until $I$ reaches a maximum. Therefore in this
  stage, in the ``competition'' between nodes in state $\mathcal{E}$ and
  $\mathcal{I}$ to turn active nodes into a new state, the external
  inactive nodes ``win''.
\item ($iii$): In this regime, the fraction of $\mathcal{I}$ nodes
  decreases while the fraction of the $\mathcal{E}$ ones is still
  growing. However since the external inactive nodes can recover more
  quickly than the internal inactive nodes ($\gamma_E>\gamma_I$) the
  probability that $k_A\leq m$ decreases. This implies that finally
  $E$ reaches a maximum and then decreases very quickly, leaving
  active nodes available to fail internally and hence $I$ grows,
  repeating again the behavior of stage ($ii$).
\end{itemize}

In the inset of Fig.~\ref{fig.ERSFesq} (b) we show, as an example, the
results of the evolution of $A$, $I$ and $E$ on a simplex triangle for
different stochastic realizations with $p^*=0.85$, which are in well
agreement with the theoretical result. In Sec.~\ref{Sec.StaAn} we will
study qualitatively this oscillatory behavior through a stability
analysis and show how this regime depends on the parameters, using a
mean field (MF) approach.

\subsection{Steady state}\label{Sec.SteStaEff}
\begin{figure}[H]
\centering
\vspace{0.5cm}
  \begin{overpic}[scale=0.30]{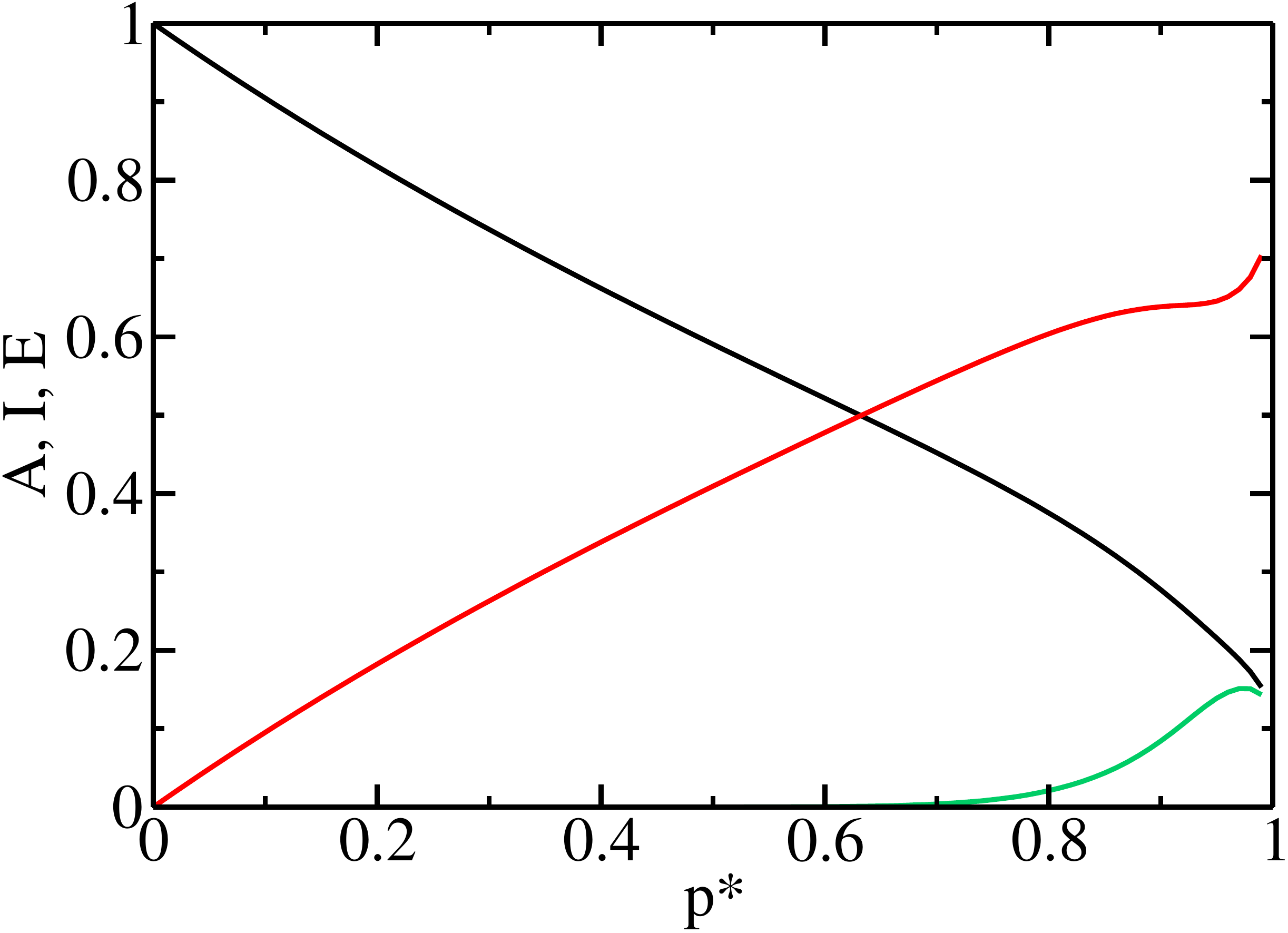}
    \put(15,50){{\bf{(a)}}}
  \end{overpic}\hspace{.50cm}
  \begin{overpic}[scale=0.30]{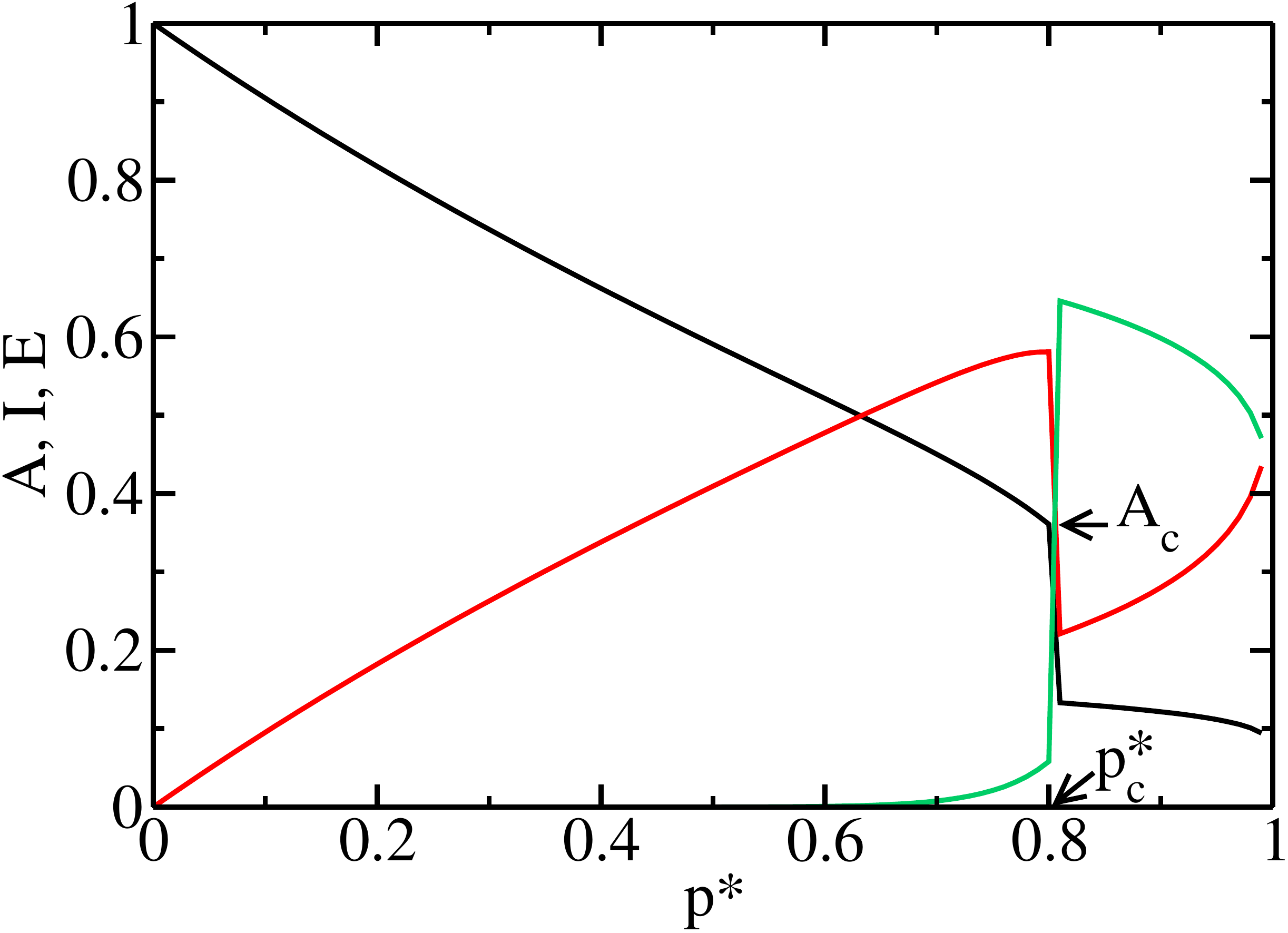}
    \put(15,50){{\bf{(b)}}}
  \end{overpic}\hspace{0.5cm}
\caption{$A$ (black), $I$ (red) and $E$ (green) as a function of
  $p^{*}$ for RR networks with $\gamma_I=0.01$, $\gamma_E=1$,
  $m=8$, and (a) $r=1$ and (b) $r=5$ obtained from the effective
  degree approximation Eqs.~(\ref{eq.lind1})-(\ref{eq.lind3}). In
  Fig.(b), $p_c^{*}=0.81$ and $A_c=0.35$ is the fraction of active
  nodes for $p^*=p_{c}^{*}-\delta p$. $A_c$ and $p_{c}^{*}$ are
  denoted by arrows.}\label{fig.TempW}
\end{figure}

Another important feature of our dynamical model is the behavior of
$A$, $I$ and $E$ in the steady state (the non oscillatory region). In
Fig.~\ref{fig.TempW}, we show the density of nodes in state
$\mathcal{A}$, $\mathcal{I}$ and $\mathcal{E}$ as a function of
$p^{*}=1-\exp(-p/\gamma_I)$ in the steady state for a random regular
network. These curves are obtained from the evolution
Eqs.~(\ref{eq.lind1})-(\ref{eq.lind3}) for $\gamma_I=0.01$,
$\gamma_E=1$, $m=8$ for different values of $r$ and initial condition
$A=1$. For $r=1$ [Fig.~\ref{fig.TempW} (a)], we can see that as the
effective rate of internal failure $p^{*}$ increases, as expected, the
density of nodes in state $\mathcal{I}$ increases while for nodes in
state $\mathcal{A}$ decreases.
In Fig.~\ref{fig.TempW} (b) we show the same curves as in Fig.  5 (a)
for $r=5$. As $p^{*}$ increases from $p^{*}=0$ the curves behaves
similar to the case $r=1$. However, at a certain value $p^* \approx
0.81$, denoted as the threshold $p^{*}_{c}$, we can see a sharp change
in the curves, like in a first order phase transition, in which the
density of nodes in states $\mathcal{A}$ and $\mathcal{I}$ abruptly
goes down, while the density of $\mathcal{E}$ grows sharply. As $p^*$
increases, for $p^{*}>p^{*}_c$ the density of nodes in state
$\mathcal{A}$ changes slower than for $p^{*}<p^{*}_c$. This implies
that the variation in the density of nodes in state $\mathcal{E}$ is
transferred to the density of nodes in state $\mathcal{I}$, $i.e.$
$\mathcal{I}$ nodes win over $\mathcal{E}$ for these parameters.

\begin{figure}[H]
\centering
\vspace{0.5cm}
  \begin{overpic}[scale=0.27]{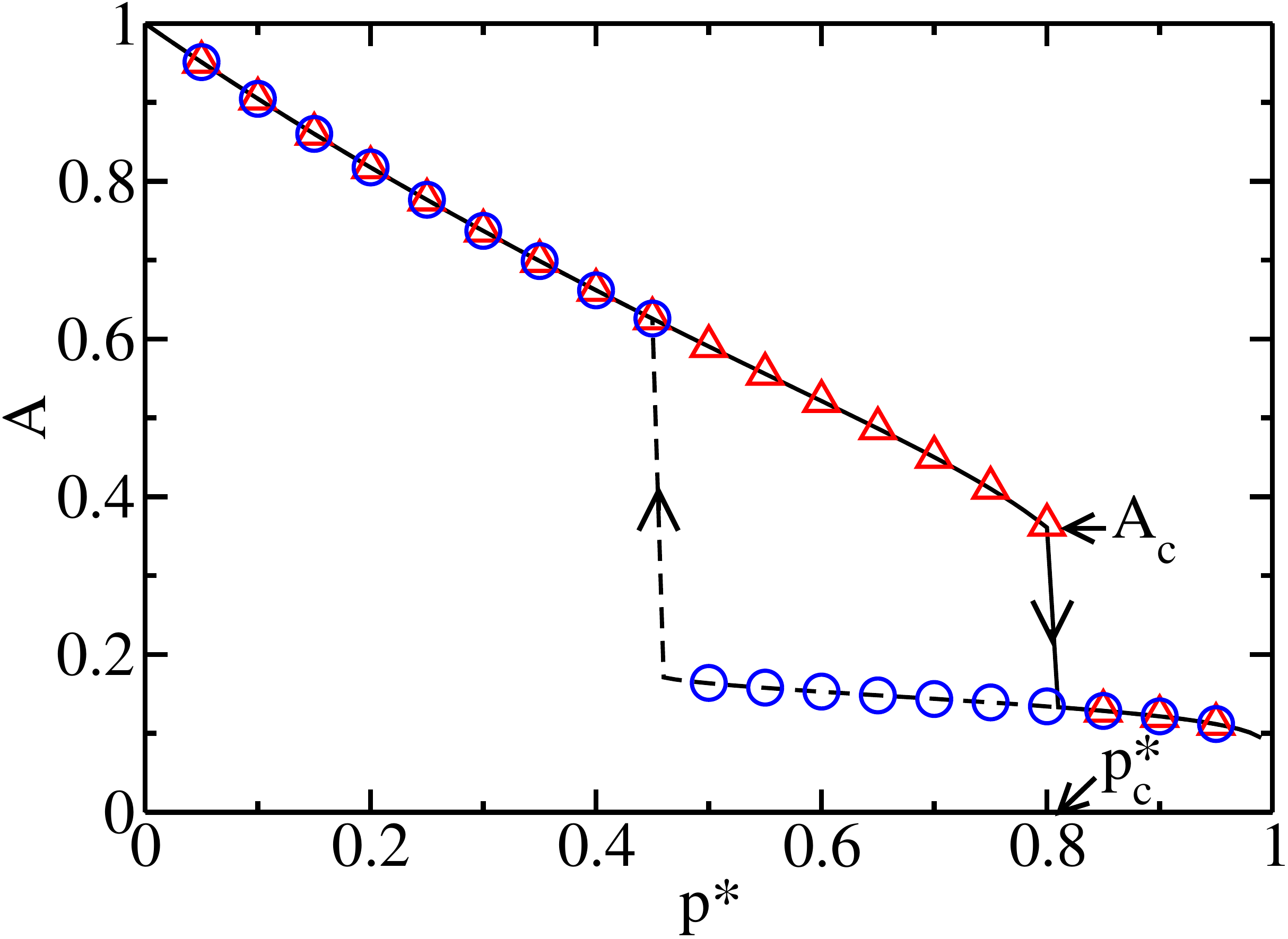}
    \put(80,60){{\bf{(a)}}}
  \end{overpic}\hspace{1.0cm}
  \begin{overpic}[scale=0.27]{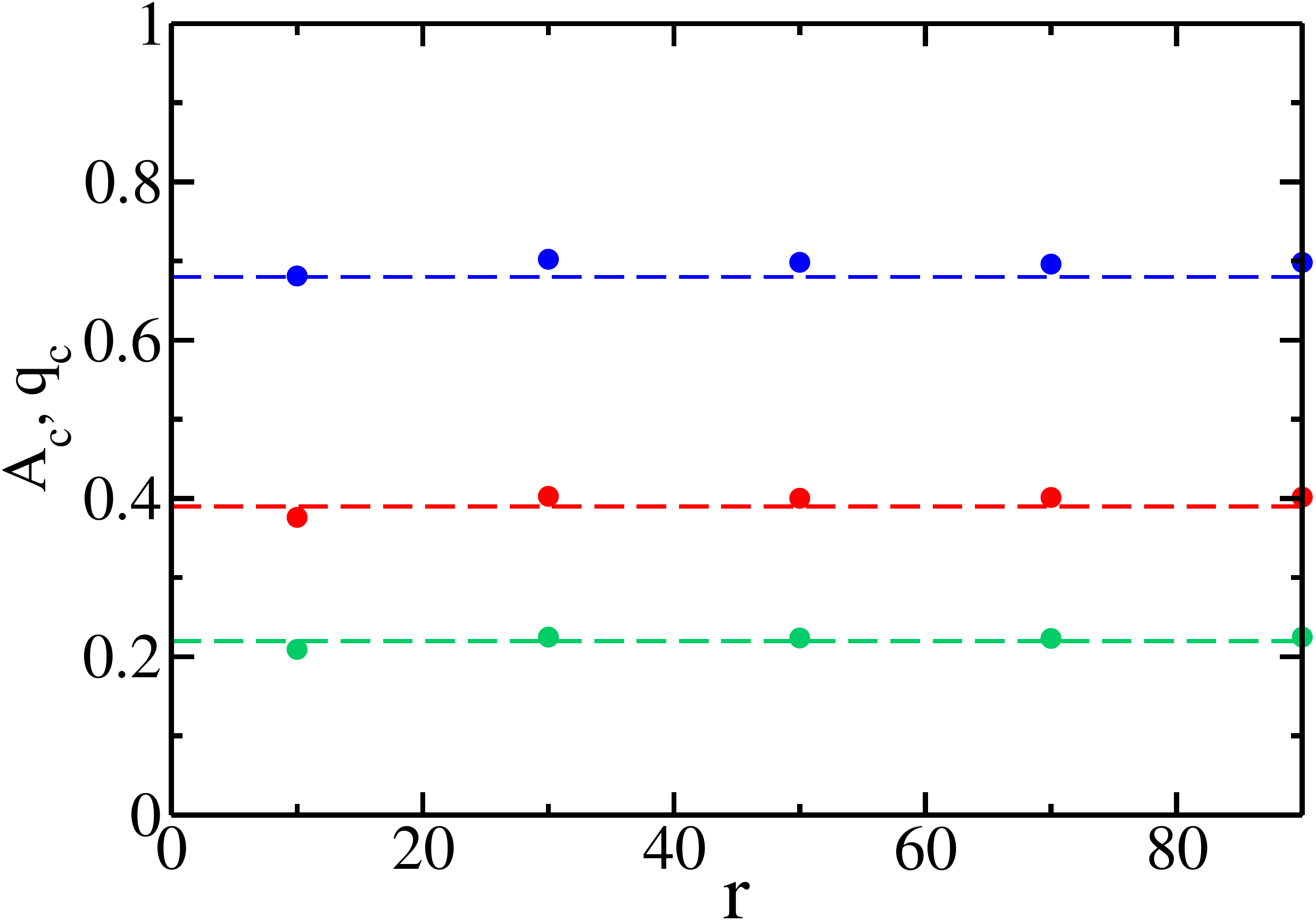}%
    \put(80,60){{\bf{(b)}}}
  \end{overpic}\hspace{1.0cm}
\caption{Figure (a): Density of active nodes in the steady state as a
  function of $p^{*}$ for RR networks with $\gamma_I=0.01$,
  $\gamma_E=1$, $m=8$ and $r=5$. The lines were obtained from the
  effective degree approach. Dashed lines (solid lines) correspond to
  the case where the initial condition consists in all nodes in state
  $\mathcal{I}$ (in state $\mathcal{A}$). The symbols correspond to
  the stochastic simulations in which the initial condition is $I=1$
  (blue circles) and $A=1$ (red triangles). The value of $A_c$ is
  denoted by an horizontal arrow. The vertical arrows indicate the
  direction of the hysteresis loop. Figure (b): The steady fraction of
  active nodes $A_c$ in our model (symbols) obtained from the
  Eqs.~(\ref{eq.lind1})-(\ref{eq.lind3}) as a function of $r$ in RR
  networks for different values of $m$: $m=4$ (green), $m=8$ (red),
  $m=16$ (blue) for $\gamma_I=10^{-2}$ and $\gamma_E=1$. We compare
  the values of $A_c$ with the critical fraction of non-removed nodes
  in ``random'' k-core percolation $q_{c}$ at which there is a first
  order transition that depends on $m$. The values of $q_c$ are
  displayed by dashed lines with the same colors as $A_c$. To compute
  the value of $A_c$ for each value of $r$, we evaluate the final
  fraction of active nodes for $p^*\in (0,1)$ with $\delta
  p^*=10^{-2}$, and then choose the value of $p^*=p^{*}_{c}$ above
  which there is a sharp decrease in $A$.}\label{fig.Stat1}
\end{figure}

In order to assess the accuracy of the theoretical approach, in
Fig.~\ref{fig.Stat1} we compare the theoretical results with the
stochastic simulations for initial conditions $A=1$ and $I=1$. We can
see a good agreement between the effective degree approach and the
simulations. For $r=5$ [see Fig.~\ref{fig.Stat1} (a)] we obtain
theoretically a hysteresis region in the density of active nodes
between $p^*=0.45$ and $p^*=0.81$. In Sec.~\ref{Sec.StaAn} we will
also study qualitatively the hysteresis through a stability analysis
in the MF approximation.

The observed sharp drop in the density of active nodes for the initial
condition $A=1$ close to $p_{c}^{*}$ [see Fig.~\ref{fig.Stat1} (a)],
in the theory and simulations, is reminiscent of the first order
transition found in ``random'' k-core percolation~\cite{Dor_01}. In the
latter process there is a critical initial fraction of removed nodes
(similar to inactive nodes in our spontaneous failure-recovery model)
that triggers a sharp decrease in the fraction of living nodes (as
mentioned in Sec.~\ref{Sec.mod} and Appendix~\ref{Sec.Akcore}).
Interestingly, we find a similitude between our model and ``random'' k-core
percolation because the value of the steady fraction of active nodes
just before the first order transition (with initial condition $A=1)$,
denoted by $A_c$, is near to the critical value of the control
parameter $q$ of this percolation process. 

In Fig.~\ref{fig.Stat1} (b) we plot the value of $A_c$ for
$p^{*}=p_c^{*}-\delta p^{*}$ ($i.e.$ just before $A$ goes down
sharply) for different values of $r$ and $m$ obtained from the
evolution Eqs.~(\ref{eq.lind1})-(\ref{eq.lind3}) and we compare them
with the threshold value $q_{c}$ in k-core percolation for the same
values of $m$ (see equations in Appendix~\ref{Sec.Akcore}). From the
figure we can see that the values of $q_{c}$ predicted by the
``random'' k-core process at which a giant component disappears are in
well agreement with the values $A_c$ obtained from our
failure-recovery model in RR networks.

\begin{figure}[H]
\centering
\vspace{0.5cm}
  \begin{overpic}[scale=0.28]{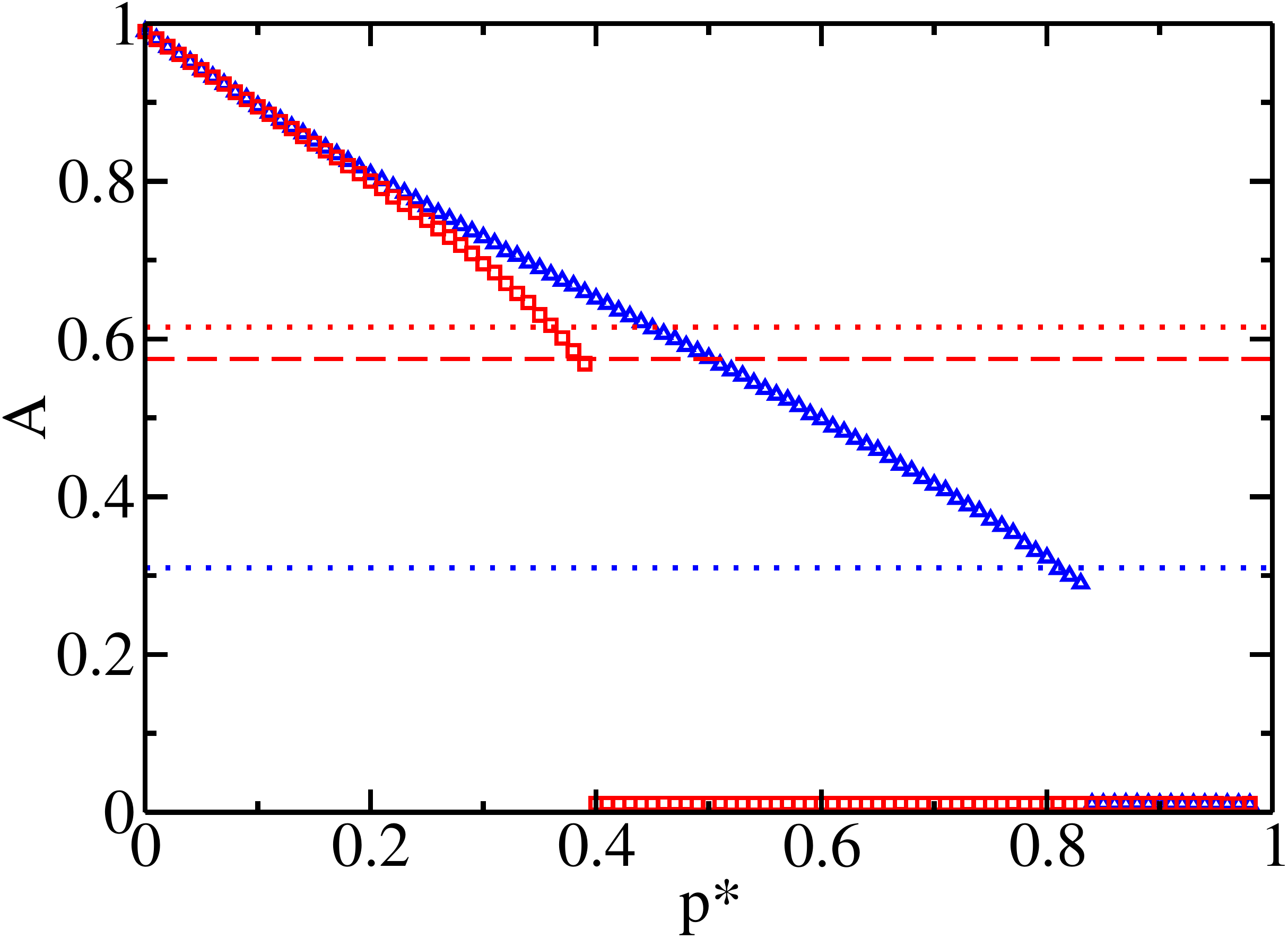}
    \put(80,60){{\bf{(a)}}}
  \end{overpic}\hspace{.50cm}
 \begin{overpic}[scale=0.28]{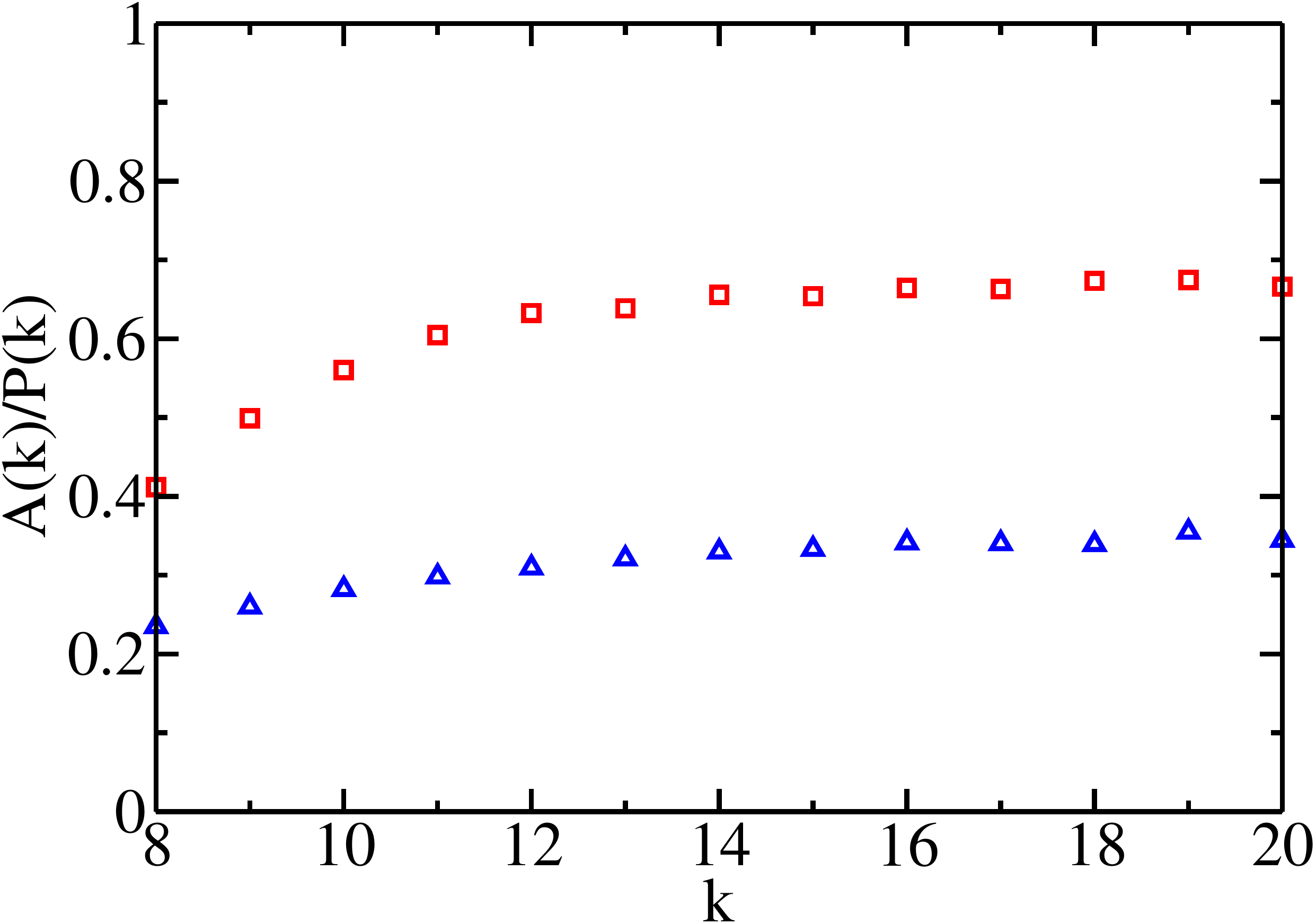}
    \put(80,60){{\bf{(b)}}}
  \end{overpic}\hspace{.50cm}
\caption{ Figure (a): Density of active nodes in the steady state as a
  function of $p^{*}$ obtained from the simulations for a network with
  a truncated Poisson distribution [see Eq.~(\ref{Eq.TPoiss})] with
  $k_{\text{min}}=8$, $k_{\text{max}}=20$ and $\lambda=10$ and for $\gamma_I=0.01$,
  $\gamma_E=1$, $r=90$, $m=2$ (blue triangles) and $m=4$ (red
  squares). The dotted lines correspond to the value of $q_c$
  predicted by the ``random'' k-core percolation, at which the fraction
  of active nodes would drop to zero if active nodes were homogeneously
  distributed. The dashed line corresponds to the value of $q_c$ obtained
  from Eqs.~(\ref{Eq.kctarg}) and~(\ref{Eq.qkq}), and using the steady
  distribution of actives nodes $q_k$ [see Eq.~(\ref{Eq.qkAct})]
  obtained just before the fall of $A$. This is explained in the end
  of Sec.~\ref{Sec.SteStaEff}. Figure (b): Steady fraction $A(k)$ of active
  nodes of connectivity $k$ relative to $P(k)$, obtained from the
  simulations just before the fall of $A$. Blue symbols correspond to
  the case $m=2$ and the red ones to $m=4$.}\label{fig.TER}
\end{figure}

In these networks the relation between the failure-recovery model and
the ``random'' k-core arises from the fact that all the nodes have the
same connectivity, and then they have the same probability to be
active. In Appendix~\ref{Sec.Arel} we explain with more detail this
relation. On the other hand, for a constant value of $m$, if we
consider the case of a broader degree distribution, such as a
truncated Poisson degree distribution
\begin{eqnarray}\label{Eq.TPoiss}
P(k) = c \frac{e^{-\lambda}\lambda^{k}}{k!}\Theta(k-k_{\text{min}})\Theta(k_{\text{max}}-k),
\end{eqnarray}
in which $c$ is a normalization constant [see Fig.~\ref{fig.TER}(a)],
we also obtain that the steady value of $A$ just before the fraction
of active nodes drops to zero is near the predicted one from
``random'' k-core percolation, in particular for lower values of $m$
. Additionally, in Fig.~\ref{fig.TER} (b) we plot the probability that
a node is active, given that it has connectivity $k$. We can see that
nodes with $k=k_{\text{min}}$ have the lowest probability to be
active. Nevertheless, the fact that for this network
$1-P(k_{\text{min}})=0.86$ and besides that the probability that a node is
active remains nearly constant disregarding its connectivity, imply
that ``random'' k-core percolation predicts approximately the value of
$A_c$. However, as we will show below, if we consider a higher
heterogeneity on the connectivities of the nodes than in the previous
case, we obtain that the steady state of the process can be better
described by a ``targeted'' k-core process rather than by the
``random'' k-core percolation.

\begin{figure}[H]
\centering
\vspace{0.5cm}
  \begin{overpic}[scale=0.28]{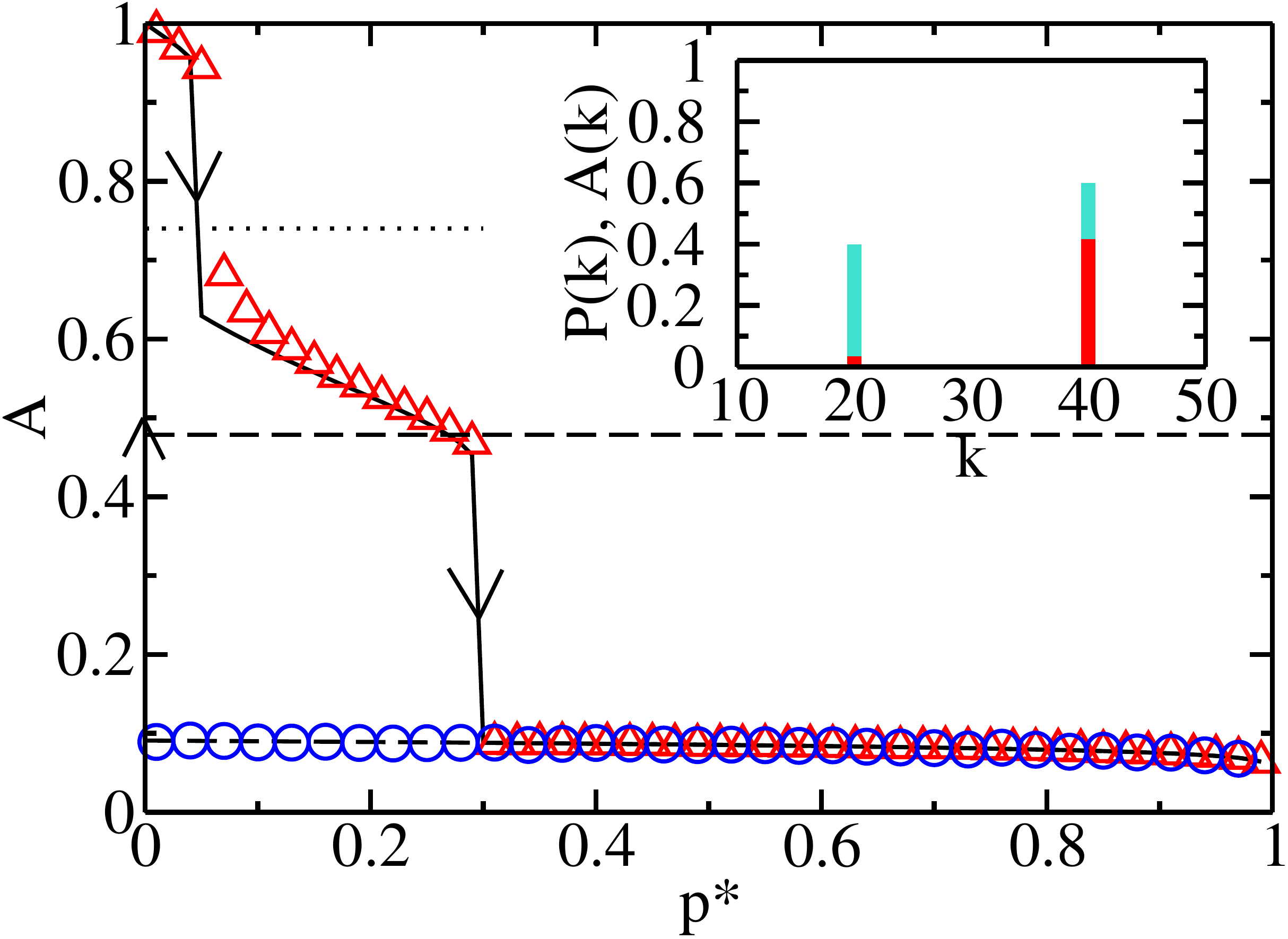}
    \put(80,60){{\bf{(a)}}}
  \end{overpic}\hspace{.50cm}
 \begin{overpic}[scale=0.28]{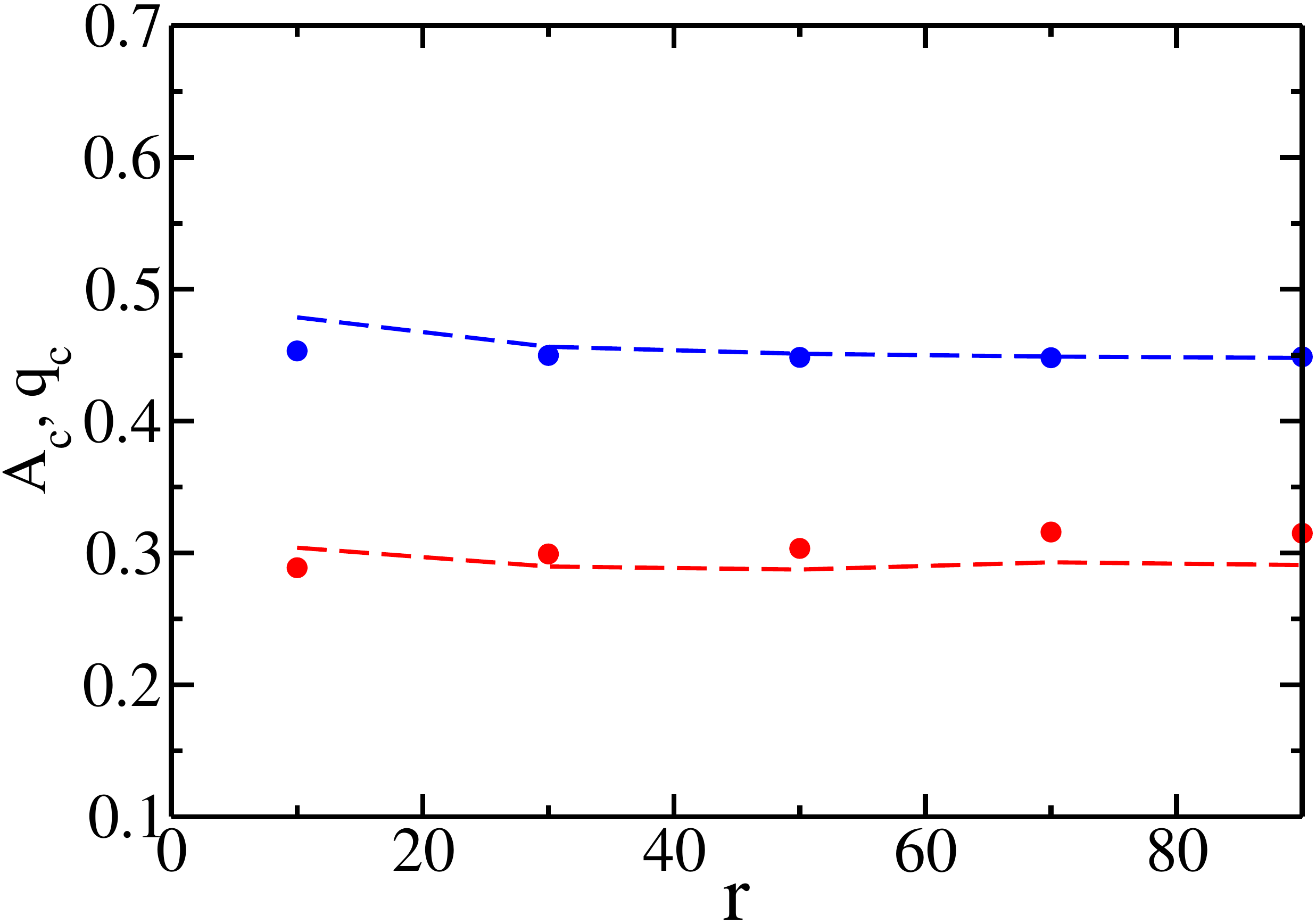}
    \put(80,60){{\bf{(b)}}}
  \end{overpic}\hspace{.50cm}
\caption{Figure (a): Density of active nodes in the steady state as a
  function of $p^{*}$ for a bimodal network with connectivities $k=20$
  and $k=40$ with $\langle k \rangle =32$ for $\gamma_I=0.01$,
  $\gamma_E=1$, $m=16$ and $r=10$. The solid lines were obtained from
  the effective degree approach. The symbols and solid lines have the
  same meaning as in Fig.~\ref{fig.Stat1}. The dotted line corresponds
  to the value of $q_c$ predicted by the ``random'' k-core
  percolation. The dashed line corresponds to the value of $q_c$ obtained
  from Eqs.~(\ref{Eq.kctarg}) and~(\ref{Eq.qkq}), and using the steady
  distribution of actives nodes $q_k$ [see Eq.~(\ref{Eq.qkAct})]
  obtained just before the second fall of $A$. In the inset we show
  the bar graphic of $P(k)$ (light blue), and $A(k)$ (red) measured in
  the steady state just before the second fall. Figure (b): The steady
  fraction of active nodes $A_c$ in our model (symbols) obtained from
  the Eqs.~(\ref{eq.lind1})-(\ref{eq.lind3}) as a function of $r$ in
  bimodal networks for different values of $m$: $m=8$ (red) and $m=16$
  (blue) for $\gamma_I=10^{-2}$ and $\gamma_E=1$. Dashed lines were
  obtained using Eqs.~(\ref{Eq.kctarg})-(\ref{Eq.qkAct}) as explained
  in the main text. }\label{fig.Bimod}
\end{figure}

In Fig.~\ref{fig.Bimod} (a) we show the steady fraction of active
nodes as a function of $p^{*}$ for a bimodal network with
connectivities $k=20$, $k=40$ and mean connectivity $\langle k
\rangle=32$ for $m=16$. From the figure we can see that for the
initial condition $A=1$, the system can exhibit two transitions for
high enough value of $m$. This is expected since as $p^{*}$ increases,
after the first sharp transition the nodes with the lowest
connectivity will fail, while the nodes with the largest connectivity
will remain active [see inset of Fig.~\ref{fig.Bimod}(a)]. Therefore,
just before the second transition ($p^*\lesssim 0.30$) the
distribution of active nodes is not homogeneous, and as a consequence
the ``random'' k- core percolation is not appropriate to describe the
steady state. For this case, in Appendix~\ref{Sec.Akchim} we present
the equations of the ``targeted'' k-core percolation that takes into
account the inhomogeneous distribution of active nodes that we will
use to compute $q_c$. In Fig.~\ref{fig.Bimod} (b) we compare $A_c$
with the value of $q_c$ obtained following Appendix~\ref{Sec.Akchim},
for a bimodal network for different values of $m$ and $r$.

From the figure we can see that the values of $q_{c}$ predicted by the
``targeted'' k-core process at which the giant component disappears
are in well agreement with the values $A_c$ obtained from our
failure-recovery model in bimodal networks. Additionally, using
``targeted'' k-core percolation, we also compute the value of $q_{c}$
for the truncated Poisson distribution [see dashed line in
  Fig.~\ref{fig.TER}(a)] with $m=4$; in which we obtain that this
value is closer to $A_c$ than the one obtained by ``random'' k-core
percolation. Therefore, these results suggest that the equations of
``targeted'' k-core percolation could be considered in non-regular
networks and used as a benchmark to compare the results with a failure-recovery
model.

In the following section we will show, using the mean field approach,
the region of parameters where the system has hysteresis and
oscillatory behaviors.

\section{Stability Analysis through mean field equations}\label{Sec.StaAn}
\subsection{Deduction of Mean Field equations}
In order to study the oscillating and hysteresis regions of our
failure-recovery spontaneous model, we use the mean field equation
(MF) derived from the effective degree approach. In particular for RR
networks, these equations depict a dynamics in which nodes shuffle
their links instantaneously~\cite{Mil_02}. While in this approach the
information about the structure of the network is lost, we can
estimate the region of parameters where the hysteresis and the
oscillatory phase exist.

Adding the system of Eqs.~(\ref{eq.lind1})-(\ref{eq.lind3}) over $k_A$, $k_I$
and $k_E$, we obtain the following equations

\begin{eqnarray}
\frac{dA}{dt}&=&\gamma_I I +\gamma_E E -r \sum_{k_A=0}^{m}\sum_{k_I=0}^{k_{\text{max}}}\sum_{k_E=0}^{k_{\text{max}}} A(k_A,k_I,k_E)-pA,\label{eqAcm1}\\
\frac{dI}{dt}&=&-\gamma_I I + p A, \label{eqIcm1} \\
\frac{dE}{dt}&=& r \sum_{k_A=0}^{m}\sum_{k_I=0}^{k_{\text{max}}}\sum_{k_E=0}^{k_{\text{max}}} A(k_A,k_I,k_E) -\gamma_E E. \label{eqEcm1}
\end{eqnarray}

Notice that these equations do not depend on $W_A$, $W_I$, $W_E$ because
the terms with these coefficients cancel each other after the addition
of the equations mentioned above.  Since $A+I+E=1$, the evolution
equations can be written as
\begin{eqnarray}
\frac{dA}{dt}&=&\gamma_I I +\gamma_E (1-A-I) -r \sum_{k_A=0}^{m}\sum_{k_I=0}^{k_{\text{max}}}\sum_{k_E=0}^{k_{\text{max}}} A(k_A,k_I,k_E)-pA,\label{eqAcm2}\\
\frac{dI}{dt}&=&-\gamma_I I + p A.
\end{eqnarray}
Using a mean field approximation, the third term of Eq.~(\ref{eqAcm2}) can be approximated by
\begin{eqnarray}
\sum_{k_A=0}^{m}\sum_{k_I=0}^{k_{\text{max}}}\sum_{k_E=0}^{k_{\text{max}}} A(k_A,k_I,k_E)=A\sum_{k=k_{\text{min}}}^{k_{\text{max}}}P(k)\sum_{k_A=0}^{m}\binom{k}{k_A}(1-A)^{k-k_A}A^{k_A},\label{eqAcm3}
\end{eqnarray}
and thus the evolution equations in the MF approach are given by
\begin{eqnarray}
\frac{dA}{dt}&=&\gamma_I I +\gamma_E (1-A-I) -r A\sum_{k=k_{\text{min}}}^{k_{\text{max}}}P(k)\sum_{k_A=0}^{m}\binom{k}{k_A}(1-A)^{k-k_A}A^{k_A} -pA,\label{eqAcm4}\\
\frac{dI}{dt}&=&-\gamma_I I + p A.\label{eqAcm5}
\end{eqnarray}

At the steady state of the process $dA/dt=dI/dt=0$, and thus $A$
satisfies the following self-consistent equation [obtained from
  Eqs.~(\ref{eqAcm4}) and ~(\ref{eqAcm5})]
\begin{eqnarray}\label{Eq.SolEquili}
A=\left(1-\frac{p}{\gamma_I}A\right) - \frac{r}{\gamma_E}A\sum_{k=k_{\text{min}}}^{k_{\text{max}}}P(k)\sum_{k_A=0}^{m}\binom{k}{k_A}(1-A)^{k-k_A}A^{k_A}.
\end{eqnarray}
Despite that in the steady state the value of $A$ depends only on the
ratios $p/\gamma_I$ and $r/\gamma_E$, the stability of the solutions
or fixed points depends on the individual values of the parameters. In
order to study the stability of the fixed points we linearize the
equations (\ref{eqAcm4}) and ~(\ref{eqAcm5}) around the fixed points
obtained from Eq.~(\ref{Eq.SolEquili}) and compute the eigenvalues of
the Jacobian matrix evaluated at the steady
state~\cite{Exxon_04}.

In the following we will analyze the stability of the solutions in the
steady state for $\gamma_I<\gamma_E$ and $\gamma_I>\gamma_E$, and show
that only for $\gamma_I<\gamma_E$, the system can sustain oscillations.

\subsection{Steady states for $\gamma_I<\gamma_E$}
\begin{figure}[H]
\centering
\vspace{1.0cm}
  \begin{overpic}[scale=0.30]{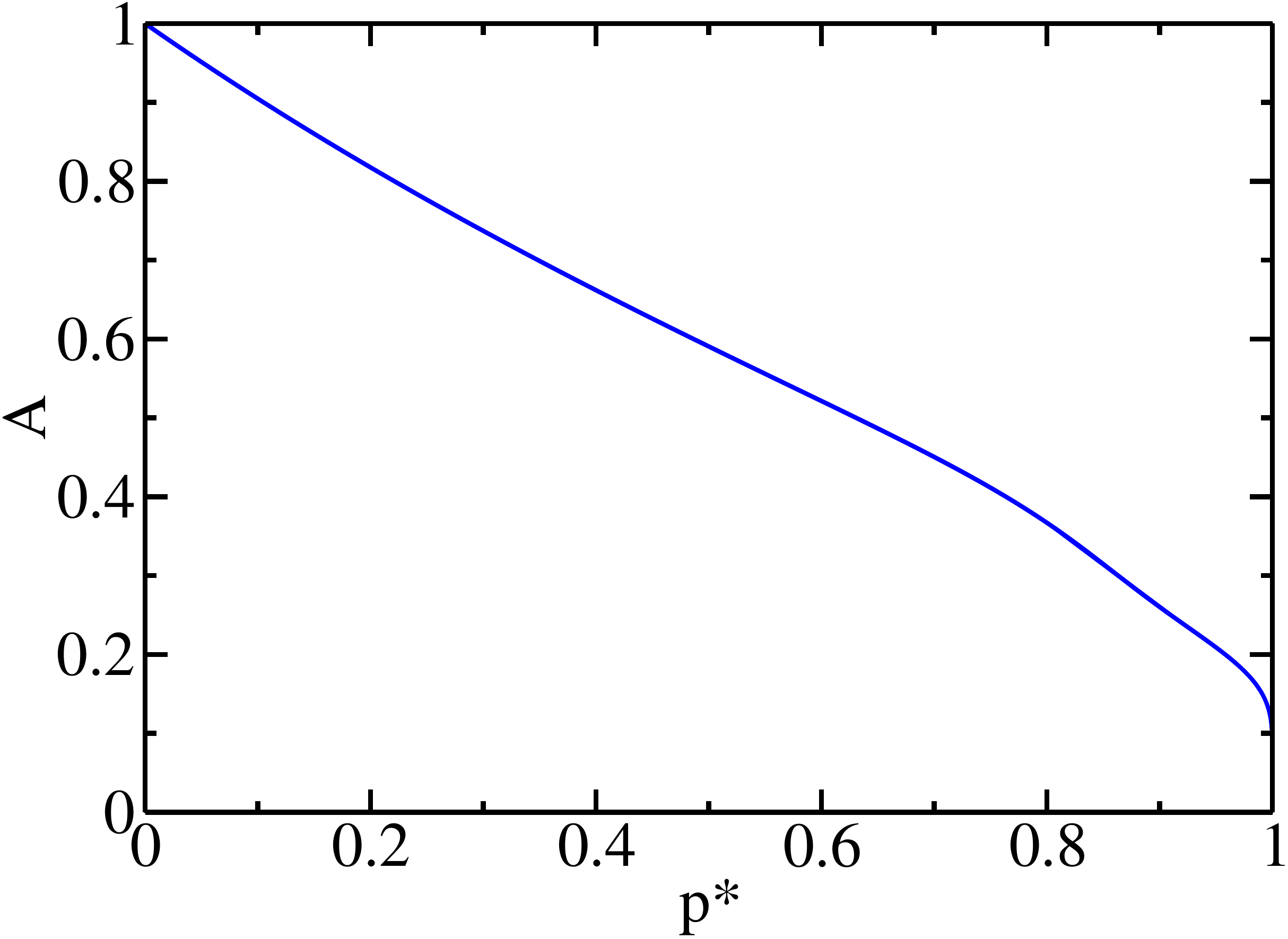}
    \put(20,30){{\bf{(a)}}}
  \end{overpic}\hspace{0.25cm}
  \begin{overpic}[scale=0.30]{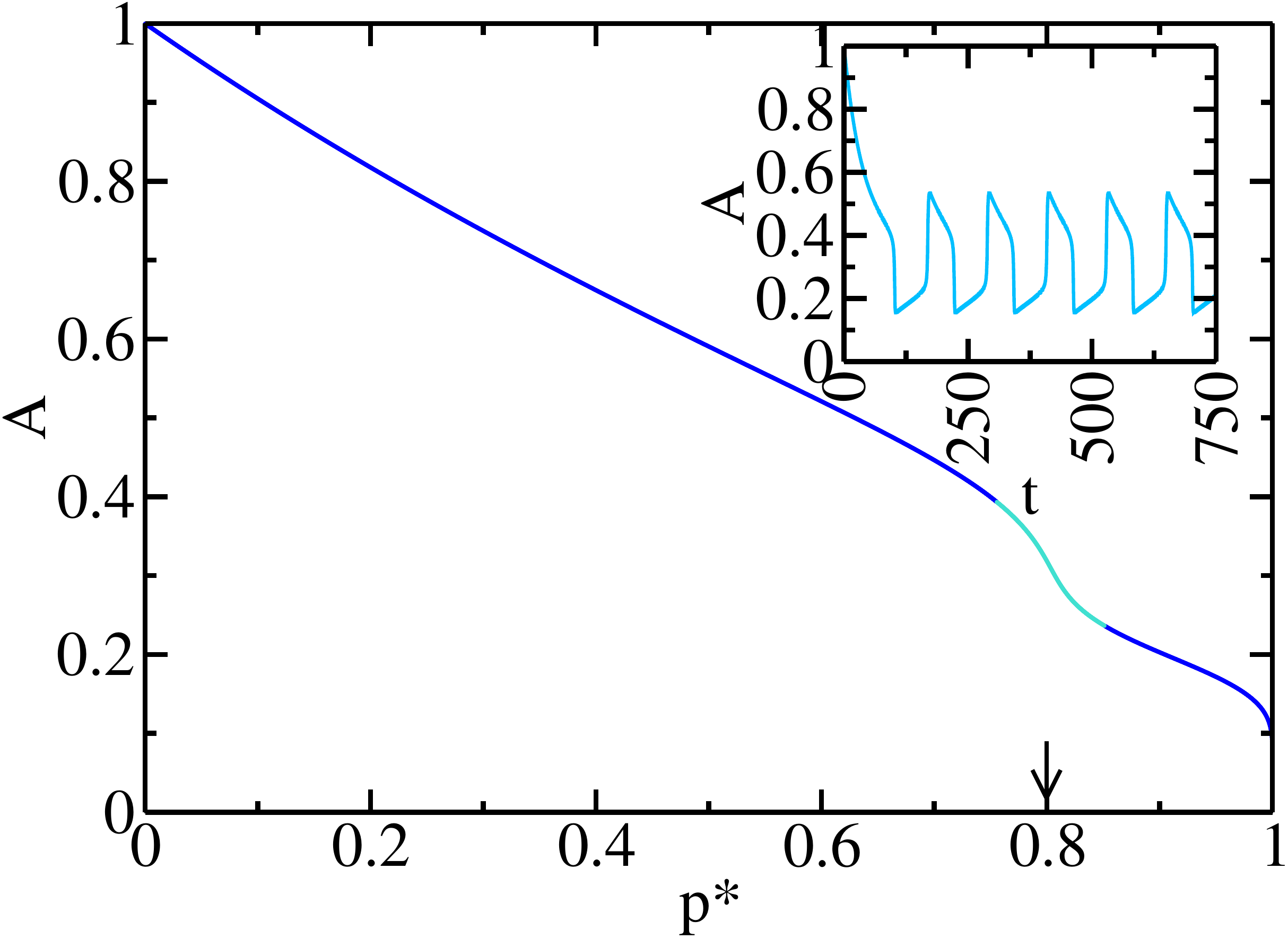}
    \put(20,30){\bf{(b)}}
  \end{overpic}\vspace{0.25cm}
  \begin{overpic}[scale=0.30]{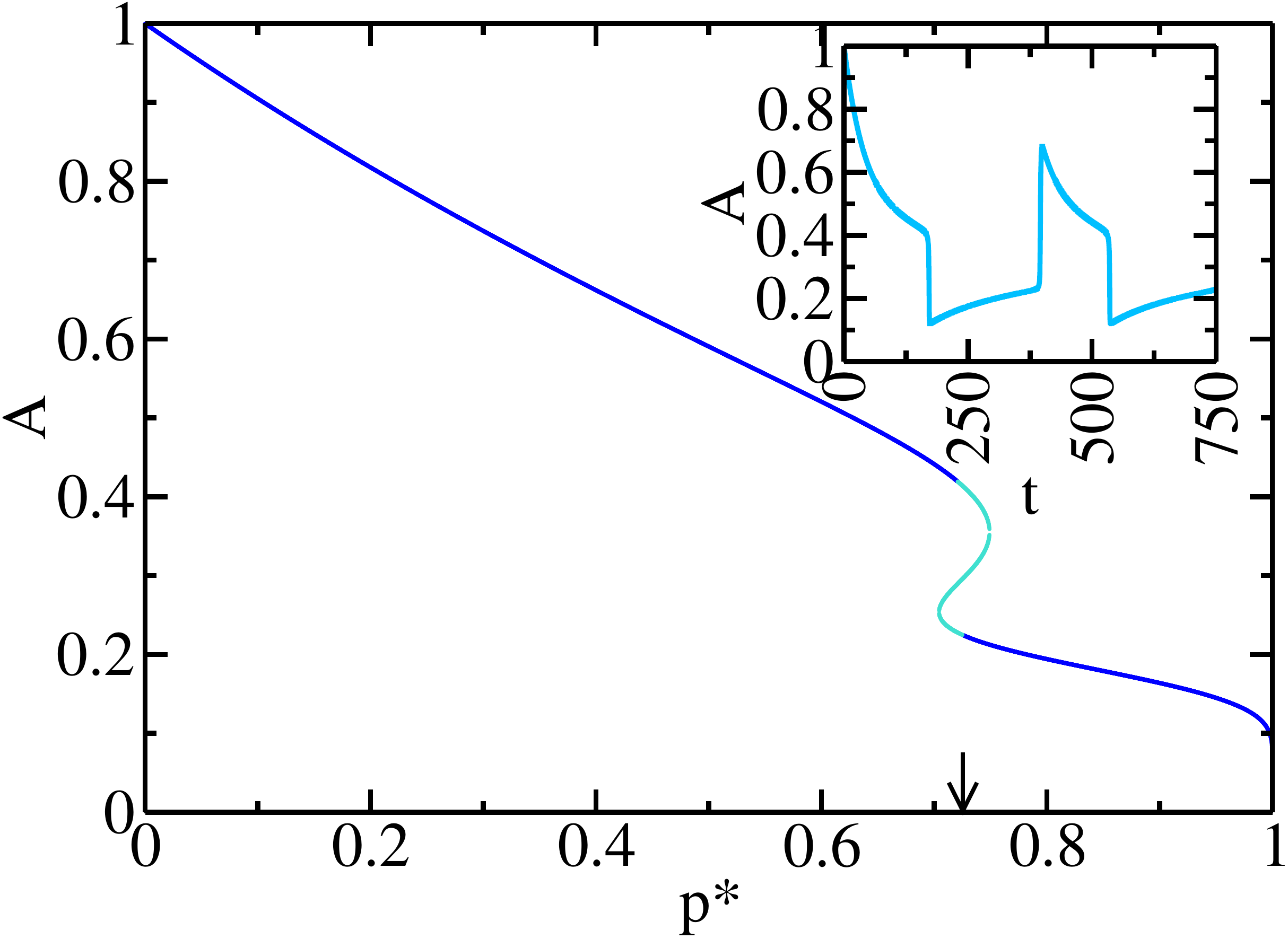}
    \put(20,30){\bf{(c)}}
  \end{overpic}\vspace{0.25cm}
  \begin{overpic}[scale=0.30]{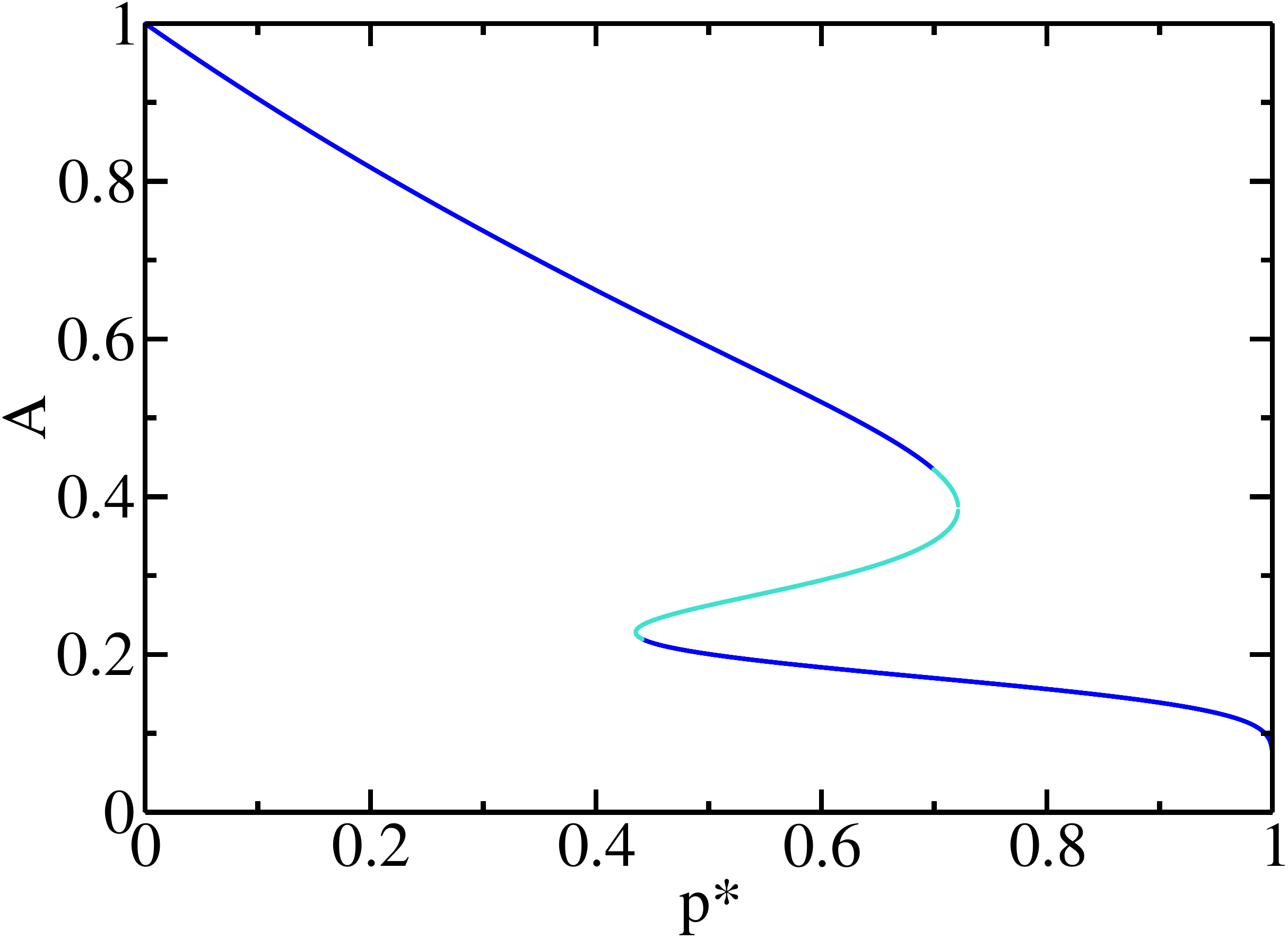}
    \put(20,30){\bf{(d)}}
  \end{overpic}\vspace{0.25cm}
\caption{Steady state of $A$ as a function of $p^{*}$ for a RR network
  with $\gamma_I=0.01$, $\gamma_E=1$, $m=8$ and $r=1$ (a), $r=2$ (b),
  $r=3$ (c) and $r=4$ (d). The curves represent the fixed points
  obtained from Eq.~(\ref{Eq.SolEquili}). Colored lines represent
  different stability-regimes obtained from the eigenvalues of the
  system of Eqs.~(\ref{eqAcm4}) and ~(\ref{eqAcm5}): light blue
  (unstable) and blue (stable). In the insets of figures $(b)$ and
  $(c)$  we show the temporal evolution of the average
  density of active nodes [obtained from Eqs.~(\ref{eqAcm4}) and
    ~(\ref{eqAcm5})] for the values of $p^{*}$ indicated by the
  arrow.}\label{Fig.Stab1}
\end{figure}

In Fig.~\ref{Fig.Stab1} we show the stability of the solutions of the
density of active nodes for different values of $r$ for RR networks
with $z=32$, $\gamma_I=0.01$, $\gamma_E=1$, $m=8$.

We observe that for $r=1$, [see Fig.~\ref{Fig.Stab1}(a)] there is only
one stable fixed point of Eq.~(\ref{Eq.SolEquili}) for each value of
$p^{*}$ . As $r$ increases to $r=2$ and $r=3$, for a range of values
of $p^{*}$ the fixed points of Eq.~(\ref{Eq.SolEquili}) are all
unstable, and therefore the densities oscillate~\cite{Exxon_05} [see
  figures ~\ref{Fig.Stab1}(b) and ~\ref{Fig.Stab1}(c)], $i.e.$ an
oscillatory regime appears for the case $\gamma_I<\gamma_E$, as
observed also in the effective degree formalism (see
Fig.~\ref{fig.ERSFesq}). Finally, for the largest value of $r$ [$r=4$,
  see figure~\ref{Fig.Stab1}(d)] a hysteresis region appears, $i.e.$
there are two stable fixed points of Eq.~(\ref{Eq.SolEquili}) for some
values of $p^{*}$.

\begin{figure}[H]
\centering
\vspace{0.5cm}
 \begin{overpic}[scale=0.30]{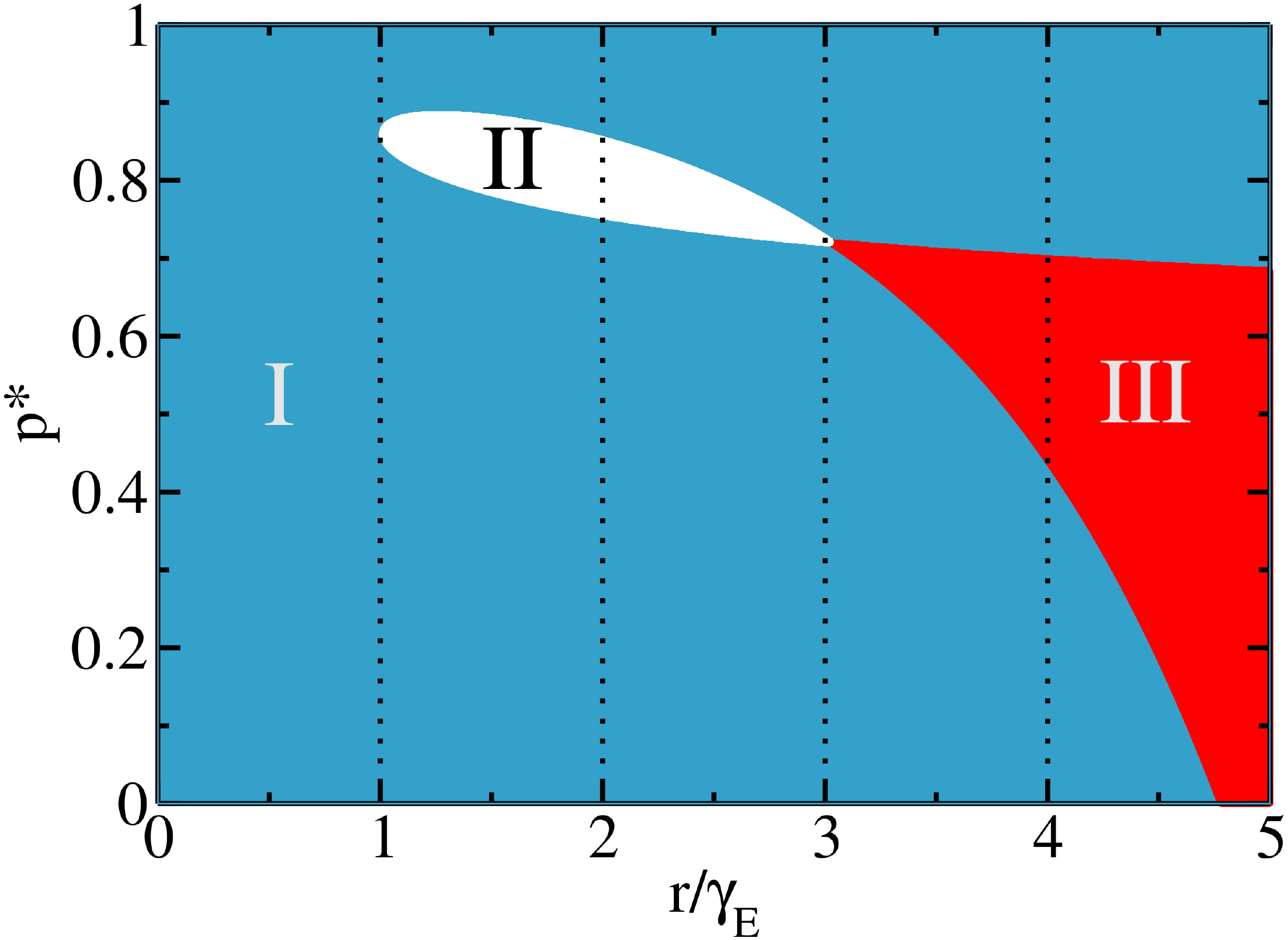}
    \put(20,30){\bf{(a)}}
  \end{overpic}\hspace{.50cm}
\vspace{0.5cm}
 \begin{overpic}[scale=0.50]{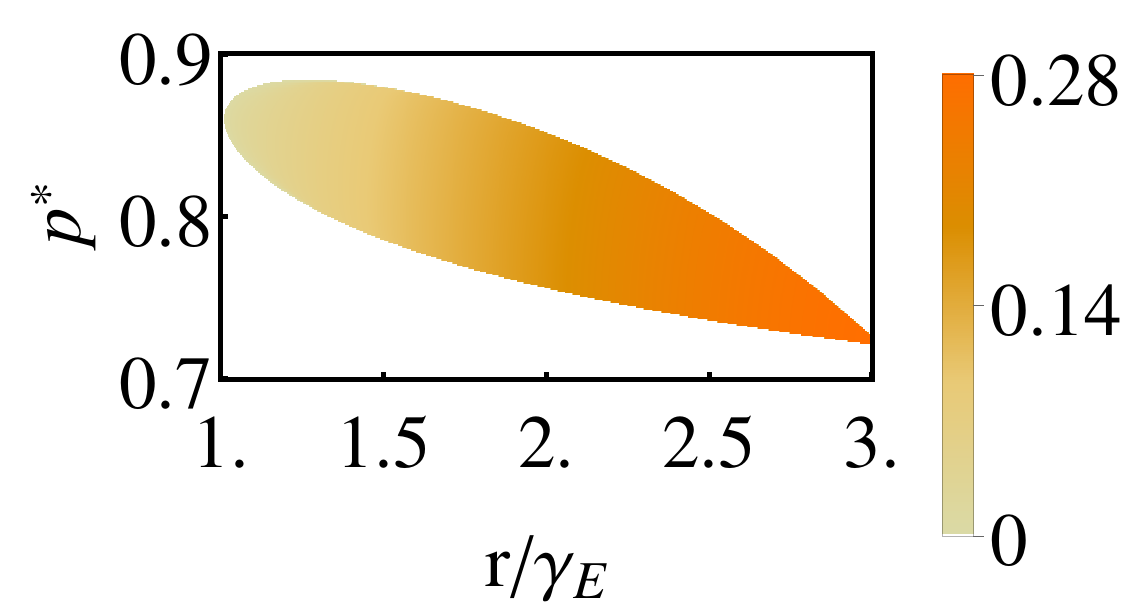}
    \put(25,25){\bf{(b)}}
  \end{overpic}\hspace{.50cm}
\vspace{0.5cm}
 \begin{overpic}[scale=0.50]{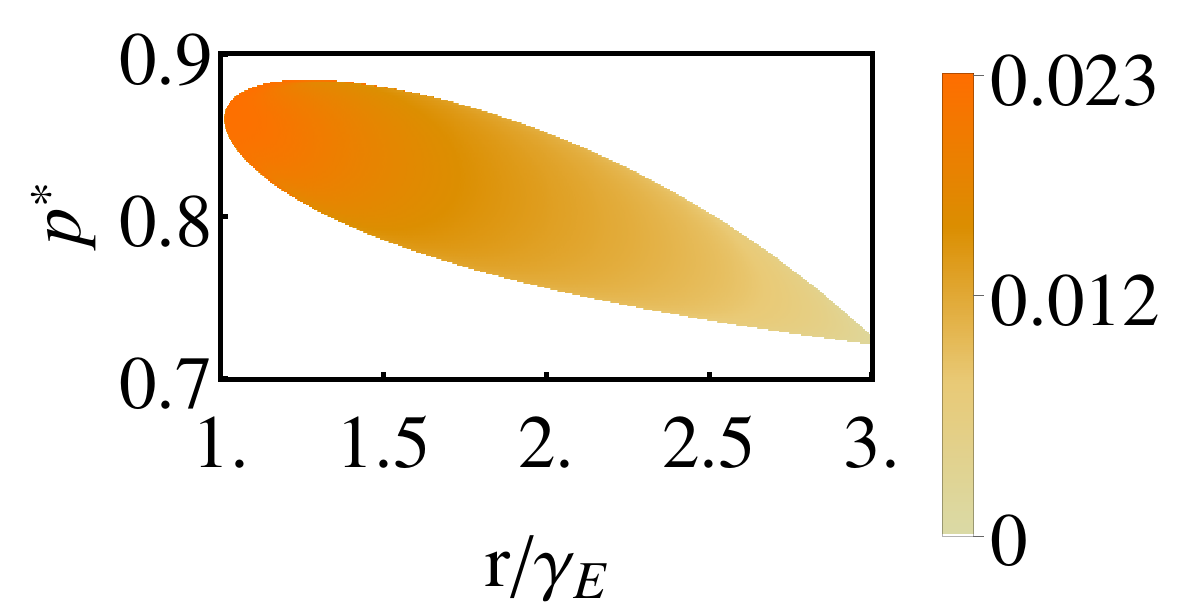}
    \put(25,25){\bf{(c)}}
  \end{overpic}\hspace{.50cm}
\caption{Figure (a): phase diagram in the plane $p^{*}$-$r/\gamma_E$
  for $\gamma_E=1$ and $\gamma_I=0.01$. The region I (blue)
  corresponds to one fixed point of the fraction of active nodes,
  region II (white) corresponds to an oscillatory regime and region
  III (red) depicts the parameters of the hysteresis region . The
  vertical dotted lines correspond to the paths on the phase diagram
  studied in Fig.~\ref{Fig.Stab1}. Figure (b): the amplitude of the
  oscillations in region II of figure (a). Figure (c): frequency
  (computed as $1/T$, where $T$ is the period) of the oscillations in
  region II of figure (a).}\label{fig.phase}
\end{figure}

Notice that the mean field equations~(\ref{eqAcm4}) and (\ref{eqAcm5})
qualitatively captures all the regimens observed in our model for the
case $\gamma_I<\gamma_E$. In Fig.~\ref{fig.phase}(a) we show the three
regimens in the plane $p^*$-$r/\gamma_E$ in which the oscillatory
region is bounded but not negligible. Therefore the oscillatory
behavior is robust in a scenario at which the parameters can vary
slightly over time within this region. This is an important fact for
biological systems in which sustained oscillations are
present~\cite{Rap_01,Gho_01,Lis_01,Men_01,Jal_01}. In order to study
the dependency of the amplitude and the frequency on the parameters,
we measure directly these magnitudes from the integration of
Eqs.~(\ref{eqAcm4}) and (\ref{eqAcm5}). In Figs.~\ref{fig.phase} (b)
and (c) we plot the amplitude and frequency respectively, which shows
that for larger values of $r/\gamma_E$ the system oscillates slower
but with a higher amplitude, which is consistent with the fact that
the parameters are close to the hysteresis region [see region III in
  Figs.~\ref{fig.phase}(a)]. Furthermore, we observe that in this
system the amplitude of the oscillations can be suppressed abruptly
when crossing the transition line, from region II to I. Finally, we
obtain that the frequency and the amplitude are more sensible under
variations in $r/\gamma_E$ than under variations of $p^*$. This result
is compatible with the shape of the oscillations, shown in
Fig.~\ref{fig.ERSFesq}(a), which have similar amplitudes and
frequencies for different values of $p^*$.

\subsection{Steady regimes for $\gamma_I>\gamma_E$}
In order to study the steady state for $\gamma_I>\gamma_E$ we will
construct the Lyapunov function $V(A,I)$ which allows to study the
global stability of a system. The Lyapunov function can only be used
if its derivative with respect to time is negative~\cite{Max_01}. In
order to achieve this goal, we next show that the dynamical
Eqs.~(\ref{eqAcm4}) and (\ref{eqAcm5}) can be expressed as a
non-gradient flow~\cite{Max_01}, $i.e.$
\begin{eqnarray}
\frac{dA}{dt}&=&-a \frac{\partial V(A,I)}{\partial A},\label{eq.AppPot1}\\
\frac{dI}{dt}&=&-b \frac{\partial V(A,I)}{\partial I},\label{eq.AppPot2}
\end{eqnarray}
where $a$ and $b$ are unknown positive constants whose values should be
consistent with Eqs.~(\ref{eqAcm4})-(\ref{eqAcm5}). Here, we
use without loss of generality, $b=1$.

\begin{figure}[H]
\centering
\vspace{0.5cm}
  \begin{overpic}[scale=0.25]{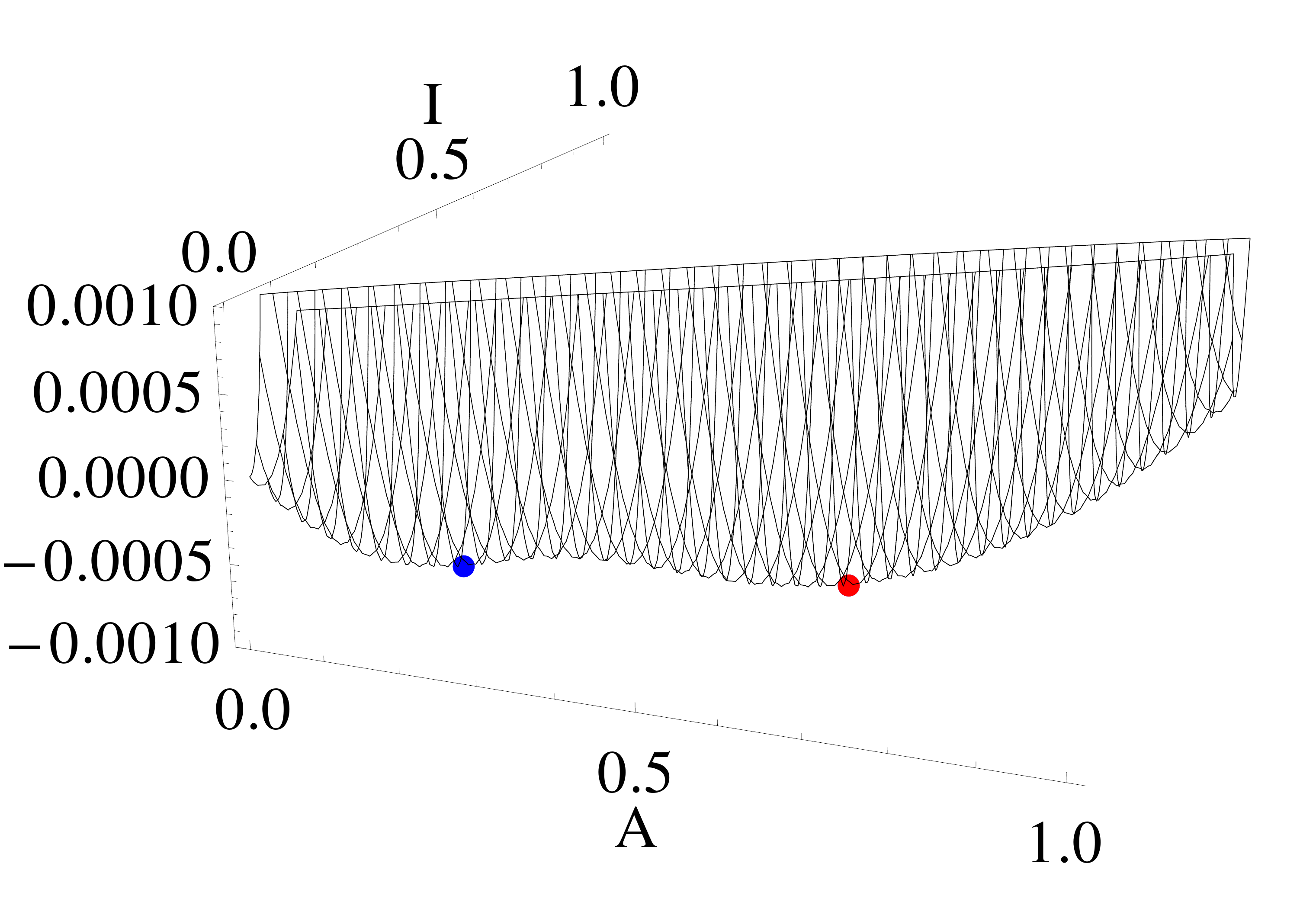}
    \put(85,60){\bf{(a)}}
  \end{overpic}\hspace{.50cm}
  \begin{overpic}[scale=0.30]{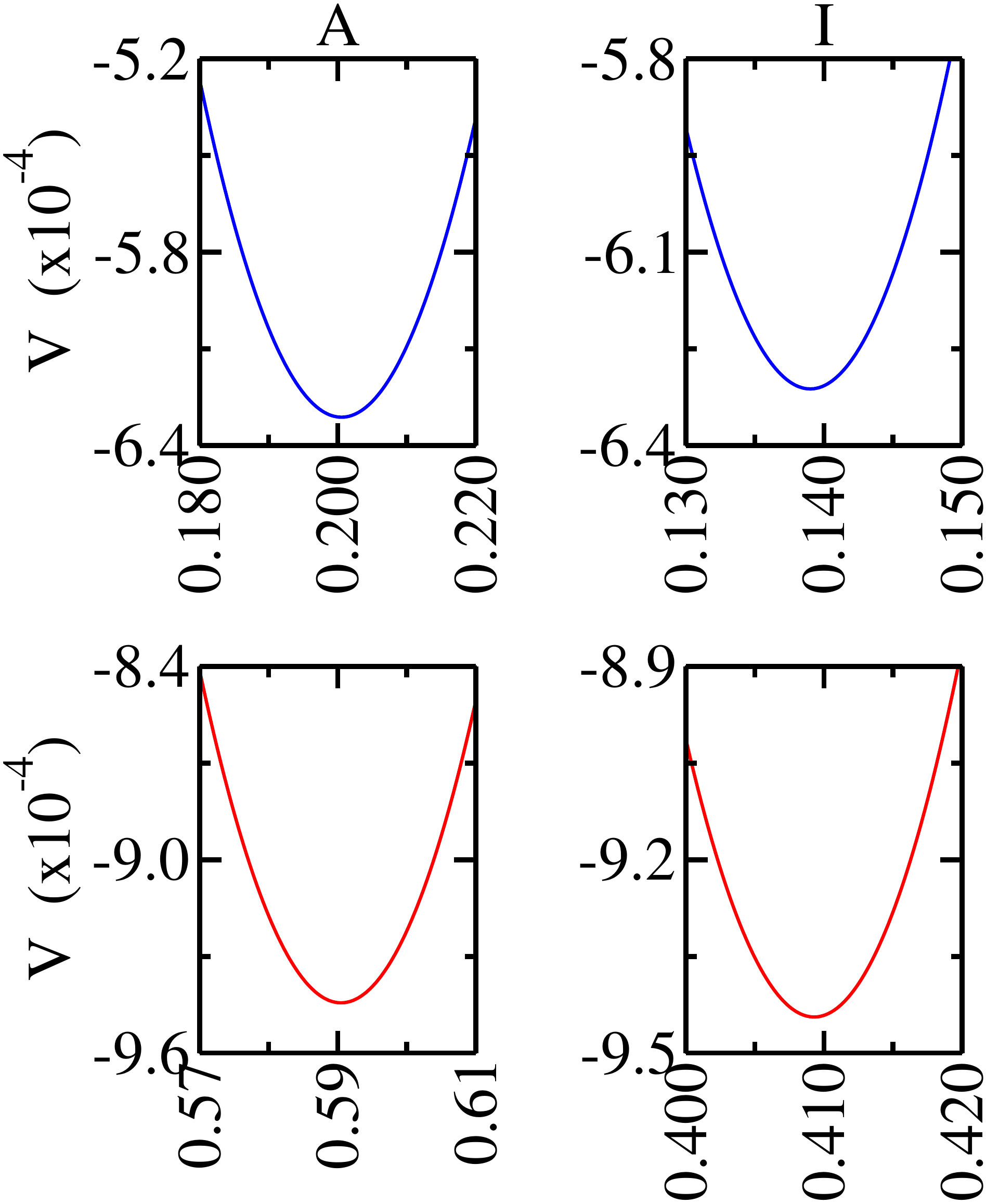}
    \put(10,100){\bf{(b)}}
  \end{overpic}\hspace{.50cm}
\caption{(a): Lyapunov function as a function of $A$ and $I$ for
  $p^{*}=0.50$ and $r=4$ for $m=8$, $\gamma_I=1$ and $\gamma_E=0.01$
  in a RR network. The red and blue points correspond to the local
  minimums of this function. Figure (b): Projection of the function
  around the local minimums of Fig.(a) in the plane $V-A$ and
  $V-I$.}\label{fig.Lyap}
\end{figure}

After matching the right hand side of
Eqs.~(\ref{eq.AppPot1})-(\ref{eq.AppPot2}) with
Eqs.~(\ref{eqAcm4})-(\ref{eqAcm5}) and integrating, the Lyapunov
function $V(A,I)$ can be written as
\begin{eqnarray}\label{Eq.AppPot}
V(A,I)=\gamma_I \frac{I^{2}}{2}-pIA+\frac{p}{\gamma_I-\gamma_E}\bigg[p\frac{A^2}{2}-\gamma_E(A-\frac{A^2}{2})+\nonumber\\r\sum_{k=k_{\text{min}}}^{k_{\text{max}}}\sum_{k_{A}=0}^{m}\sum_{j=0}^{k-k_A}P(k)\binom{k}{k_A}\binom{k-k_A}{j}(-1)^{j}\frac{A^{k_A+j+2}}{k_A+j+2}\bigg],
\end{eqnarray}
where
\begin{eqnarray}
a&=&\frac{\gamma_I-\gamma_E}{p},
\end{eqnarray}
with $\gamma_I>\gamma_E$ (in order to ensure that $a>0$). Using the
proposed Lyapunov function and the values of $a$ and $b$, it is
straightforward to show that these values allows to reconstruct
Eqs.~(\ref{eqAcm4})-(\ref{eqAcm5}) through
Eqs.~(\ref{eq.AppPot1})-(\ref{eq.AppPot2}). In Fig.~\ref{fig.Lyap}, we
plot the Lyapunov function for $\gamma_E=0.01$ and $\gamma_I=1$. From
the plot we can see that the Lyapunov function has two local minimums.

For $\gamma_I>\gamma_E$, $dV(A,I)/dt<0$, and therefore the overall
system tends to a local minimum. In addition, since $V(A,I)$ is
expressed in terms of powers of $A$ and $I$, it has a finite number of
local and isolated minimums, hence an oscillatory behavior is not
allowed because the system get stuck in a local minimum, from which it
cannot escape due to the lack of fluctuations. Notice that in the case
$\gamma_I<\gamma_E$, we cannot use the Lyapunov function given by
Eq.~(\ref{Eq.AppPot}), since for this case the parameter $a$ in
Eq.~(\ref{eq.AppPot1}) is negative. This implies that we cannot
guarantee that $dV(A,I)/dt<0$, and therefore the dynamics of the
system is not necessarily in a minimum of the function $V(A,I)$.

\begin{figure}[H]
\centering
\vspace{0.5cm}
  \begin{overpic}[scale=0.30]{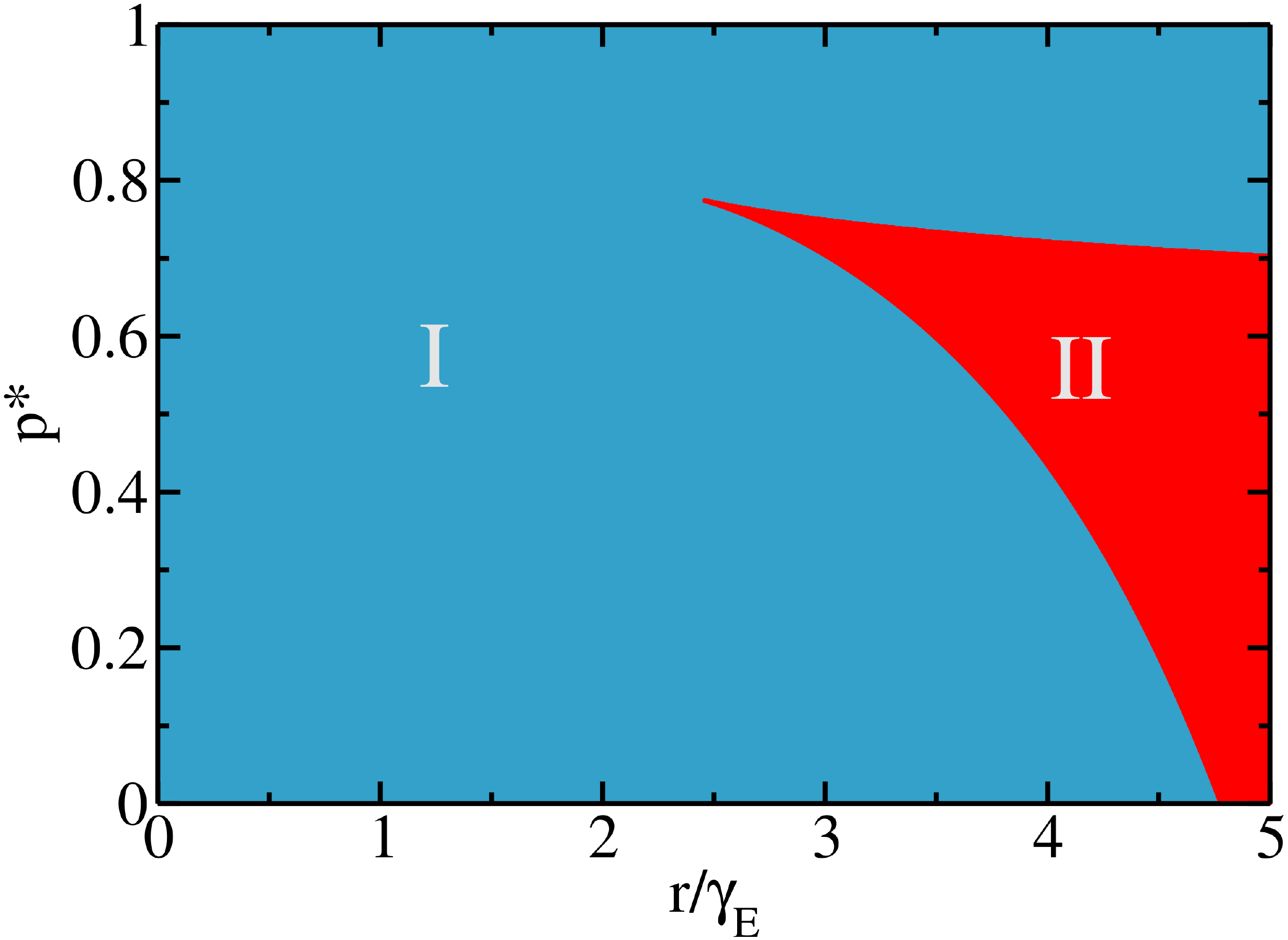}
    \put(20,60){}
  \end{overpic}\hspace{.50cm}
\caption{Phase diagram $p^{*}$ vs. $r$ for $m=8$, $\gamma_I=1$ and
  $\gamma_E=0.01$ in a RR network. Region I (blue) corresponds to
  the existence of a single value of the fraction of active nodes in
  the steady state and region II (red) depicts the parameters of the
  hysteresis region.}\label{fig.phase1fin}
\end{figure}

In Fig.~\ref{fig.phase1fin} we show the phase diagram in the plane
$p^{*}-r/\gamma_E$ for the case $\gamma_I>\gamma_E$, obtained from
Eq.~(\ref{Eq.SolEquili}). We can see that in region I there is only
one stable fixed point, while in region II there are two stable
fixed points, $i.e.$, the hysteresis behavior is present, especially
for large values of $r/\gamma_E$. However, there is not an oscillatory
regime, which is compatible with the existence of a Lyapunov
function. Therefore, the relation between $\gamma_E$ and $\gamma_I$ is
a key factor for the existence of sustained oscillations but not for
the hysteresis.

\section{Conclusion}\label{Concl}

In summary, in this work we study a failure-recovery model in which
the failure state belongs to two different kinds: internal and
external failed nodes. Using the degree effective approach and
simulations we found theoretically and via stochastic simulations that
the system may exhibit hysteresis on the fraction of active nodes and
also an oscillatory behavior as a result of the competition between
external and internal inactive nodes. In the steady state we find that
in random regular networks, the critical fraction of active nodes
below which there is an abrupt collapse is close to the threshold in
the ``random'' k-core percolation. However for non-regular networks,
the topology can lead to an inhomogeneous distribution of active nodes
which can be better described by ``targeted'' k-core percolation
rather than by a ``random'' k-core percolation. Using a MF approach,
we obtain that for $\gamma_E>\gamma_I$ there is a range of the
parameters at which the system can exhibit sustained oscillations, and
that their amplitude increases and their frequency decreases as the
parameters approach to the region at which hysteresis is
present. Finally we show through the Lyapunov function that for
$\gamma_I>\gamma_E$, the oscillatory phase is absent but can still
exist a hysteresis region.  We believe that the model we proposed and
the equations developed in this work can be the useful for future
research on dynamical systems and their relation with percolation
theory. A possible extension of our model would be to generalize our
equations to take into account heterogeneous values of $m$. Another
possible extension could be to model the process in interacting
networks~\cite{Maj_02} which could allow to understand how the
transitions can be affected by the interaction.

\appendix
\section{Derivation of $p^*$}\label{appPstar}
In this section, we obtain the parameter $p^*$ as the steady fraction
of inactive internal nodes when $\gamma_E=r=0$, which corresponds to
the case in which the nodes on the network can only be in states $\mathcal{A}$ and
$\mathcal{I}$. 

For the case $\gamma_E=r=0$ the nodes activate and fail
intermittently without interaction between them and therefore, the
temporal evolution of the fraction of nodes in state $\mathcal{A}$ and
$\mathcal{I}$ is governed by the following equations,
\begin{eqnarray}
\frac{dA}{dt}&=&\gamma_I I-p\;A,\\
\frac{dI}{dt}&=&-\gamma_I I+p\;A.\label{EqpEst2}
\end{eqnarray}
Notice that we are assuming as initial condition of the
dynamics that there are no externally failed nodes. Since $E\equiv 0$, then
$I+A=1$ and the Eq.~(\ref{EqpEst2}) reduces to,
\begin{eqnarray}
\frac{dI}{dt}&=&-\gamma_I I+p\;(1-I),
\end{eqnarray}
whose solution in the steady state is given by,
\begin{eqnarray}
I(t\to \infty) &=&\frac{p}{\gamma_I+p}.
\end{eqnarray}
For small values of $p$, the last expression can be rewritten as,
\begin{eqnarray}
I(t\to \infty) \approx \frac{p}{\gamma_I}\approx 1-\exp(-\frac{p}{\gamma_I})\equiv p^*.
\end{eqnarray}

\section{k-core Percolation}
\subsection{``Random'' k-core Percolation}\label{Sec.Akcore}

Random $k$-core percolation is an irreversible dynamical process
in which a node can be removed (dead) or non-removed (living). In the
initial state, all nodes are living and then a randomly fraction $1-q$
of nodes is removed. Afterwards, all the living nodes with $m$ or less
living neighbors, are removed. This step is repeated iteratively until
the system is composed only by living nodes with more than $m$ living
neighbors. In Ref.~\cite{Dor_01}, using a generating function
formalism, the steady state of the final fraction of living nodes in
complex networks $P_{\infty}$, was described by solving the following
self-consistent equation
\begin{eqnarray}
Q_{\infty}&=&1-q+q\sum_{k=k_{\text{min}}}^{k_{\text{max}}}\frac{kP(k)}{\langle k \rangle}\sum_{u=0}^{m-1}\binom{k-1}{u}Q_{\infty}^{k-1-u}(1-Q_{\infty})^u,\label{Eq.A01}
\end{eqnarray}
where $Q_{\infty}$ is the probability of reaching a dead node through
a randomly chosen link. The value of $Q_{\infty}$ that depends on $q$
is found solving the self-consistent equation~(\ref{Eq.A01}) in
$Q_{\infty}$.

With the solution of $Q_{\infty}$ for a given value of $q$, we obtain
the fraction of nodes in the giant component $P_{\infty}$:
\begin{eqnarray}
P_{\infty}&=&q \sum_{k=k_{\text{min}}}^{k_{\text{max}}}P(k)\left(1-\sum_{u=0}^{m}\binom{k}{u}Q_{\infty}^{k-u}(1-Q_{\infty})^{u}\right).
\end{eqnarray}

\subsection{Relation between the failure-recovery model in RR networks and ``random'' k-core percolation}\label{Sec.Arel}
\begin{figure}[H]
\centering
\vspace{0.5cm}
  \begin{overpic}[scale=0.35]{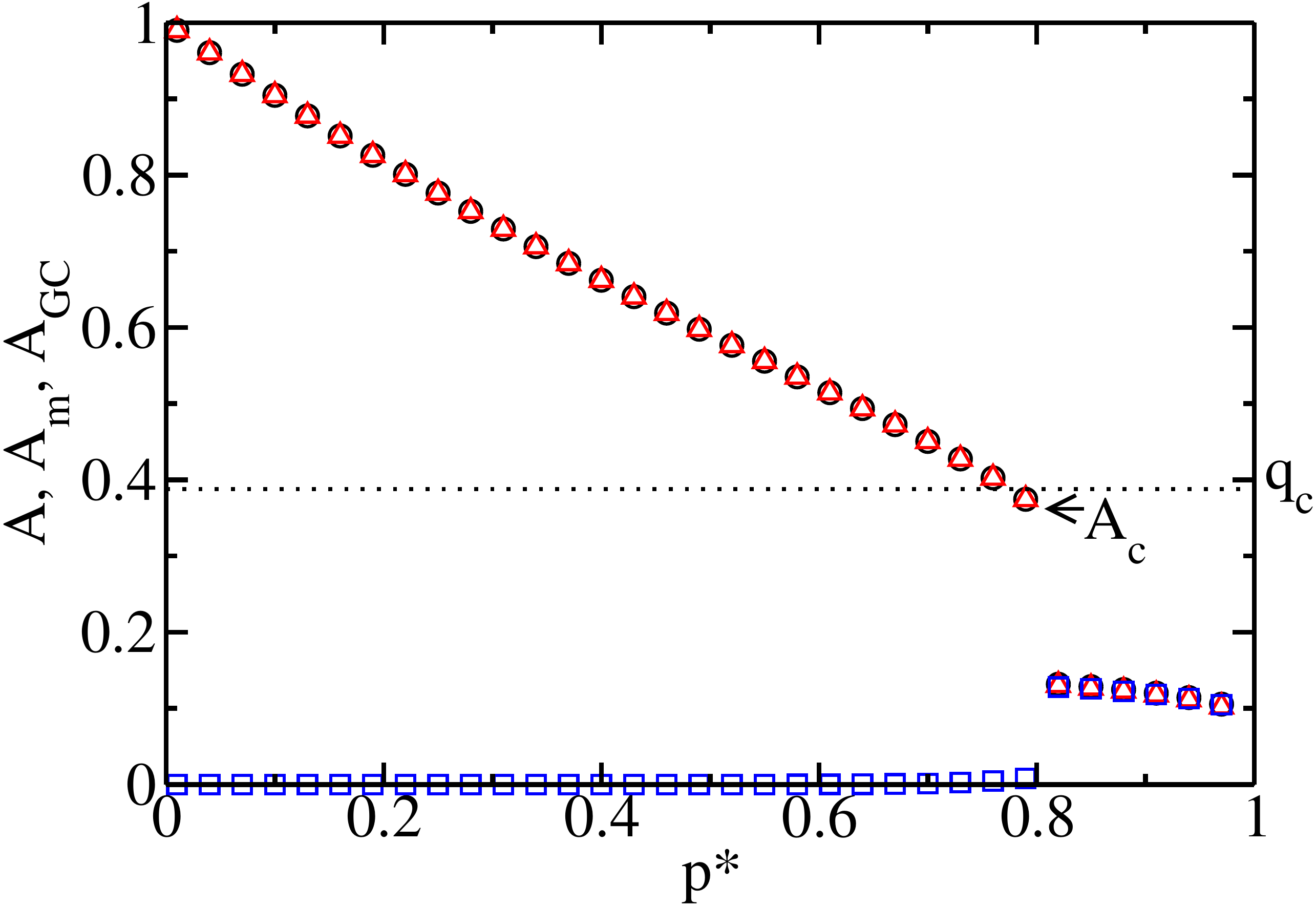}
  \end{overpic}\hspace{.50cm}
\caption{$A$ (black, $\bigcirc$), $A_{m}$ (blue, $\square$) and $A_{GC}$
  (red, $\triangle$) as a function of $p^*$ obtained from simulations for
  $m=8$, $r=10$, $\gamma_I=10^{-2}$ and $\gamma_E=1$ with
  $N=10^5$. The dotted line indicates $q_{c}=0.38$ and the arrow
  indicates $A_c=0.36$.}\label{fig.kcoreApp}
\end{figure}

In order to explain the similitude between $A_c$ and $q_{c}$ for RR
networks, discussed in Sec.~\ref{Sec.SteStaEff} [see
  Fig.~\ref{fig.Stat1} (b)], in Fig.~\ref{fig.kcoreApp} we plot the
simulations for: (i) $A$, (ii) the fraction of active nodes with $k_A
\leq m$ ($A_{m }$) and (iii) the fraction of active nodes that belong
to the GC ($A_{GC}$) as a function of $p^{*}$.
From the figure, we can see that for $A>A_c$ almost all active nodes
belong to the GC with $k_A>m$ while for $A<A_c$ almost all active
nodes have $k_A \leq m$. Heuristically, in a k-core percolation
framework, these results can be interpreted in the following way:
assuming that the total fraction of active nodes at which $A\sim
A_c\cong q_{c}$ are placed randomly on the network, k-core percolation
predicts the existence of a GC with active nodes with at least $k_A>m$
neighbors, which avoids the collapse of the system. If the fraction of
active nodes is below $A_c$ this GC with active nodes with $k_A>m$
does not exist. Therefore, if the system has a large value of $r$
($i.e.$ if the rate at which $\mathcal{A}$ goes to $\mathcal{E}$ is
large compared to the rate of recovery $\gamma_E$), then the fraction
of external inactive nodes rises sharply and $A$ collapses. Therefore
for large values of $r$, k-core percolation theory allows to estimate
approximately the value of active nodes below which there is a first
order transition.

\subsection{Targeted k-core percolation}\label{Sec.Akchim}
Given a network with degree distribution $P(k)$, let $1-q_k$ be the
probability that a node with degree $k$ is initially removed on the
cascade of failure in a k-core percolation process. Then, following
Ref.~\cite{Cal_01}, it is straightforward to show that the final fraction of
non-removed nodes is obtained solving the following equations,
\begin{eqnarray}\label{Eq.kctarg}
Q_{\infty}&=&\sum_{k=k_{\text{min}}}^{k_{\text{max}}}\frac{kP(k)}{\langle k \rangle}\left(1-q_k+q_k\sum_{u=0}^{m-1}\binom{k-1}{u}Q_{\infty}^{k-1-u}(1-Q_{\infty})^u\right), \\
P_{\infty}&=&\sum_{k=k_{\text{min}}}^{k_{\text{max}}}P(k)q_k\left(1-\sum_{u=0}^{m}\binom{k}{u}Q_{\infty}^{k-u}(1-Q_{\infty})^{u}\right),
\end{eqnarray}
where $Q_{\infty}$ is the probability of reaching a removed node
through a link. In this ``targeted'' k-core percolation process,
likewise as in the ``random'' k-core percolation, we also called $q$
the total initial fraction of non-removed nodes, $i.e.$
\begin{eqnarray}\label{Eq.qkq}
q=\sum_{k=k_{\text{min}}}^{k_{\text{max}}}P(k)q_k.
\end{eqnarray}

\begin{figure}[H]
\centering
\vspace{0.5cm}
  \begin{overpic}[scale=0.75]{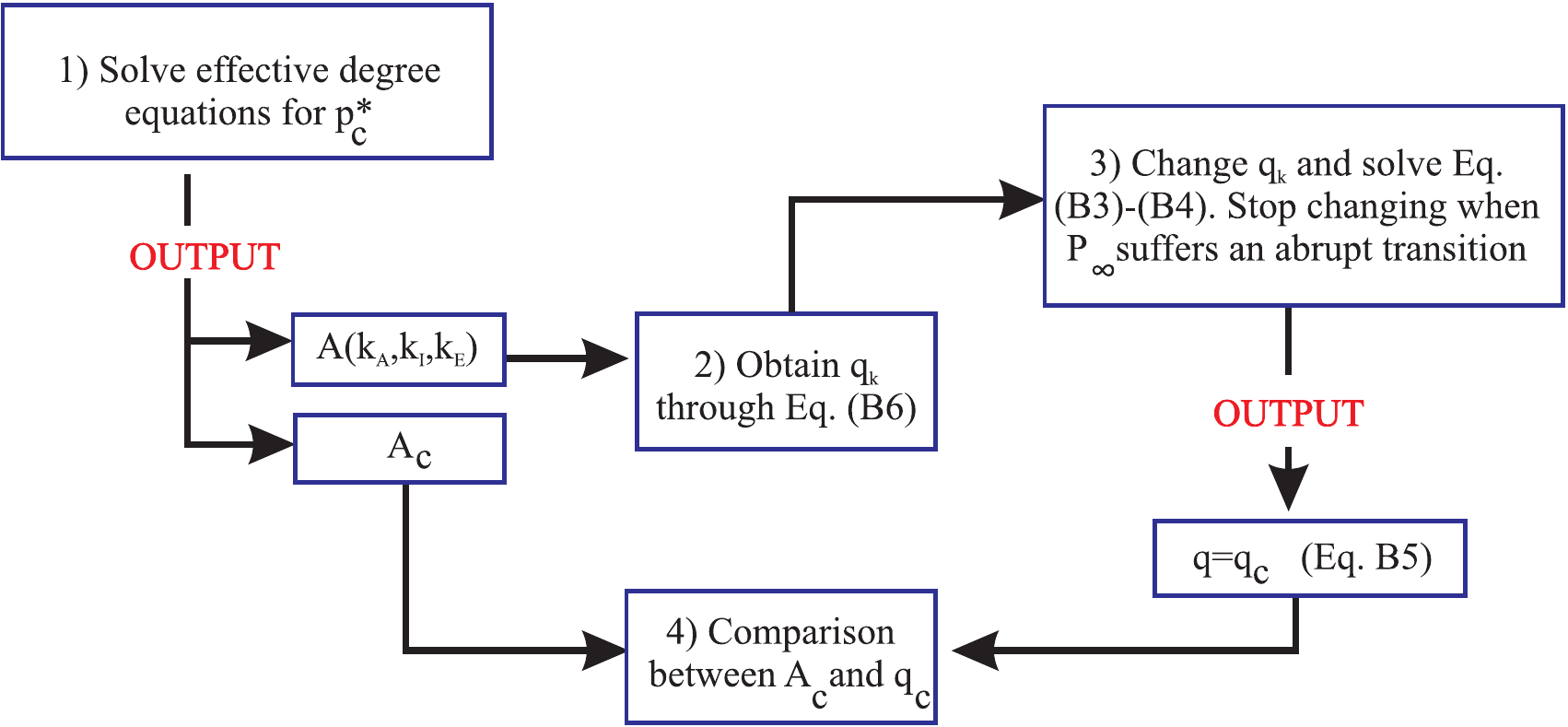}
  \end{overpic}\hspace{.50cm}
\caption{Flow diagram to compute  $q_c$ in targeted k-core
  percolation}\label{fig.Aflux}
\end{figure}

In the ``random'' k-core percolation process, a variation in the value
of $q$ implies that the fraction of non-removed nodes varies in the
same proportion independent of its connectivity. However in the
``targeted'' k-core percolation process, there is not a unique way to
change the fraction $q_k$. Therefore we propose that
for a given distribution of $q_k$ which satisfies Eq.~(\ref{Eq.qkq}),
a decreasing on the value of $q$ implies that the distribution $q_k$
decreases from its tail, $i.e.$ the non-removed nodes with the highest
connectivity are removed. A similar process is performed when the
value of $q$ is increased. Then we propose that in the steady state of
our failure-recovery model for heterogeneous degree distributions,
$q_k$ is given by
\begin{eqnarray}\label{Eq.qkAct}
q_k=\sum_{k_A=k_{\text{min}}}^{k_{\text{max}}}\sum_{k_I=k_{\text{min}}}^{k_{\text{max}}}\sum_{k_E=k_{\text{min}}}^{k_{\text{max}}}A(k_A,k_I,k_E)\delta_{k,k_A+k_I+k_E}
\end{eqnarray}
In order to show that for $A=A_c$, this distribution of non-removed
nodes is near a transition point in a ``targeted'' percolation
process, we vary the value of $q$ [given by Eq.~(\ref{Eq.qkq})]
starting from the distribution $q_k$ [see Eq.~(\ref{Eq.qkAct})] as
explained above, in order to compute $q_c$.  In Fig.~\ref{fig.Aflux}
we summarize with a schematic, the steps to compute the value of
$q_c$.

\section*{Acknowledgments}
We wish to thank to UNMdP, FONCyT and CONICET (Pict 0429/2013, Pict
1407/2014 and PIP 00443/2014) for financial support. We also thank
Dr. G. G. Iz\'us and Dr. H. H. Arag\~ao R\^ego for useful discussions.

\bibliography{bib}

\end{document}